\documentclass[english]{iopart}
\usepackage{xcolor}
\usepackage{amstext}
\usepackage{amssymb}
\usepackage{graphicx}
\PassOptionsToPackage{normalem}{ulem}
\usepackage{ulem}

\makeatletter

\providecommand{\tabularnewline}{\\}
\providecolor{lyxadded}{rgb}{0,0,1}
\providecolor{lyxdeleted}{rgb}{1,0,0}
\DeclareRobustCommand{\mklyxadded}[1]{\bgroup\color{lyxadded}{}#1\egroup}
\DeclareRobustCommand{\mklyxdeleted}[1]{\bgroup\color{lyxdeleted}\mklyxsout{#1}\egroup}
\DeclareRobustCommand{\mklyxsout}[1]{\ifx\\#1\else\sout{#1}\fi}

\usepackage{iopams}
\usepackage{setstack}

\makeatother

\usepackage{babel}
\begin{document}
\title{Expulsion of runaway electrons using ECRH in the TCV tokamak}
\author{J. Decker$^{1}$, M. Hoppe$^{2}$, U. Sheikh$^{1}$, B.P. Duval$^{1}$,
G. Papp$^{3}$, L. Simons$^{1}$, T. Wijkamp$^{4}$, J. Cazabonne$^{1}$,
S. Coda$^{1}$, E.~Devlaminck$^{1}$, O. Ficker$^{5}$, R. Hellinga$^{6}$,
U. Kumar$^{1}$, Y.~Savoye-Peysson$^{7}$, L. Porte$^{1}$, C. Reux$^{7}$,
C. Sommariva$^{1}$, A. Tema Biwol\'e$^{1}$, B. Vincent$^{1}$, L.
Votta$^{2}$, the TCV Team$^{8}$, and the EUROfusion Tokamak Exploitation
Team$^{9}$}

\address{$^{1}$Swiss Plasma Center (SPC), Ecole Polytechnique F\'ed\'erale de
Lausanne (EPFL) CH-1015 Lausanne, Switzerland}

\address{$^{2}$Department of Electrical Engineering, KTH Royal Institute
of Technology, Stockholm, Sweden}

\address{$^{3}$Max Planck Institute for Plasma Physics D-85748 Garching,
Germany}

\address{$^{4}$FOM Institute DIFFER \textquoteleft Dutch Institute for Fundamental
Energy Research\textquoteright{} 5600 HH Eindhoven, Netherlands}
\address{$^{5}$Institute of Plasma Physics of the CAS CZ-18200 Praha 8, Czech
Republic}
\address{$^{6}$\textup{Department of Applied Physics and Science Education,
Eindhoven University of Technology, Eindhoven 5600 MB, Netherlands}}
\address{$^{7}$CEA-IRFM F-13108 Saint-Paul-les-Durance, France}
\address{$^{8}$See author list of \textquotedblleft Experimental research
on the TCV tokamak\textquotedblright , by B.P. Duval, et al., to be
published in Nuclear Fusion Special Issue: Overview and Summary Papers
from the 29th Fusion Energy Conference (London, UK, 16-21 October
2023).}
\address{$^{9}$See the author list of \textquotedblleft Progress on an exhaust
solution for a reactor using EUROfusion multi-machines capabilities\textquotedblright{}
by E. Joffrin et al., to be published in Nuclear Fusion Special Issue:
Overview and Summary Papers from the 29th Fusion Energy Conference
(London, UK, 16-21 October 2023).}
\begin{abstract}
Runaway electrons (REs) are a concern for tokamak fusion reactors
from discharge startup to termination. A sudden localized loss of
a multi-megaampere RE beam can inflict severe damage to the first
wall. Should a disruption occur, the existence of a RE seed may play
a significant role in the formation of a RE beam and the magnitude
of its current. The application of central electron cyclotron resonance
heating (ECRH) in the Tokamak \`a Configuration Variable (TCV) reduces
an existing RE seed population by up to three orders of magnitude
within only a few hundred milliseconds. Applying ECRH before a disruption
can also prevent the formation of a post-disruption RE beam in TCV
where it would otherwise be expected. The RE expulsion rate and consequent
RE current reduction are found to increase with applied ECRH power.
Whereas central ECRH is effective in expelling REs, off-axis ECRH
has a comparatively limited effect. A simple 0-D model for the evolution
of the RE population is presented that explains how the effective
ECRH-induced RE expulsion results from the combined effects of increased
electron temperature and enhanced RE transport.
\end{abstract}
\maketitle

\section{Introduction}

In tokamaks, runaway electrons (REs) can be generated when the toroidal
electric field exceeds a critical value that is proportional to the
plasma density \cite{Breizman_2019}. In recent years, REs have been
studied extensively as they present a threat for reactor-scale tokamaks
where disruptions may generate multi-MA RE beams that could inflict
significant damage to the first wall \cite{LEHNEN201539}. Strong
RE populations have been observed in all phases of tokamak discharges
: plasma startup, current rampup, full current regime, and post-disruption.
In reactor-scale tokamak disruptions, RE generation is expected to
be dominated by avalanches of knock-on collisions \cite{rosenbluth97theory,Boozer_2017},
which depends critically upon the strength of a pre-disruption RE
seed \cite{Hesslow_2019}. Tokamak operation can thus be made safer
by supressing any RE seed before an eventual disruption, or by, at
least, reducing the RE density.

The Tokamak \`a Configuration Variable (TCV) is equipped with diagnostics
dedicated to fast electron physics and has developed scenarios with
strong RE currents without significant danger to the machine vessel
\cite{Decker_2022,Sheikh_2024}. With extensive real-time control
including magnetic shaping, high power auxiliary heating and multiple
gas injection systems, TCV has become particularly well suited to
RE research \cite{dec23}. In TCV, sustained RE beams are observed
following disruptions induced by massive gas injection (MGI) only
when a significant fraction of the plasma current is driven by REs
before the disruption \cite{Decker_2022,hop23}.

Runaway electrons are generated in a plasma with electron density
$n$ if the electric field $E$ parallel to the magnetic field lines
exceeds a critical value $E_{c}$ beyond which the electric force
overcomes collisional friction force of electrons that are travelling
close to the speed of light. Connor \& Hastie expressed this critical
field as \cite{connor75relativistic} 
\begin{equation}
E_{c}=\frac{ne^{3}\ln\Lambda}{4\pi\varepsilon_{0}^{2}mc^{2}},\label{eq:ec}
\end{equation}
where $\ln\Lambda$ is the Coulomb logarithm, $e$ the elementary
charge, $\varepsilon_{0}$ the free space permittivity, $m$ the electron
rest mass and $c$ the speed of light in vacuum. For $E>E_{c}$, electrons
with sufficient momentum $p>p_{c}=mc(E/E_{c}-1)^{-1/2}$ accelerate
continually and are termed runaway electrons. REs can be generated
by collisional diffusion through the $p=p_{c}$ boundary beyond which
electrons undergo this net acceleration, referred to as the Dreicer
mechanism \cite{dreicer59electron}. In addition, a knock-on collision
between an existing RE and a slower, free or bound, electron can result
in both electrons now satisfying $p>p_{c}$, termed the avalanche
process \cite{rosenbluth97theory}. Experimentally, REs are detected
only when $E\gg E_{c}$ \cite{granetz14ecrit}. By accounting for
additional physical effects such as radiation losses, radial transport,
and collisions with partially ionized impurities, \cite{ale15,Decker_2016,Tinguely_2018,Hesslow_2018},
an effective critical field larger than $E_{c}$ can be defined to
identify the conditions for RE generation \cite{mar10,sta15}.

In TCV, an observable RE population is generated in plasmas where
the density is sufficiently low such that the parallel electric field
satisfies $E/E_{c}\gtrsim10$ \cite{dec23}. At sufficiently low plasma
densities, strong RE growth is observed with the discharge entering
a regime where a significant fraction of the toroidal plasma current
is carried by REs. After an MGI-triggered disruption, a so-called
``RE beam'' carries the entire plasma current, as shown in Section
\ref{sec:expulsion}. Under otherwise identical conditions, applying
ECRH before MGI results in a near-complete expulsion of the RE seed
population preventing the formation of a RE beam following the triggered
disruption. ECRH-induced RE expulsion is further characterised in
Section \ref{sec:power}. The RE expulsion rate and the resulting
RE current reduction are found to increase with ECRH power with estimations
of RE loss rates and RE population reduction derived from experimental
observations. A radial scan of ECRH power deposition shows that the
RE population reduction is considerably stronger for centrally compared
to off-axis deposited ECRH. A 0-D analysis of the ECRH expulsion dynamics
is presented in Section \ref{sec:regimes}. The RE current decay during
ECRH is associated with a transition from high to low RE current regime
from the combined effect of higher RE losses and reduced avalanche
RE generation due to the higher bulk electron temperature and correspondingly
lower loop voltage. The paper concludes by discussing applications
of the ECRH-induced RE expulsion and proposing scenarios to further
characterise the underlying physical mechanisms.

\section{Prevention of post-disruption RE beam formation using ECRH in TCV\protect\label{sec:expulsion}}

\begin{figure}[h]
\begin{centering}
\includegraphics[width=8cm]{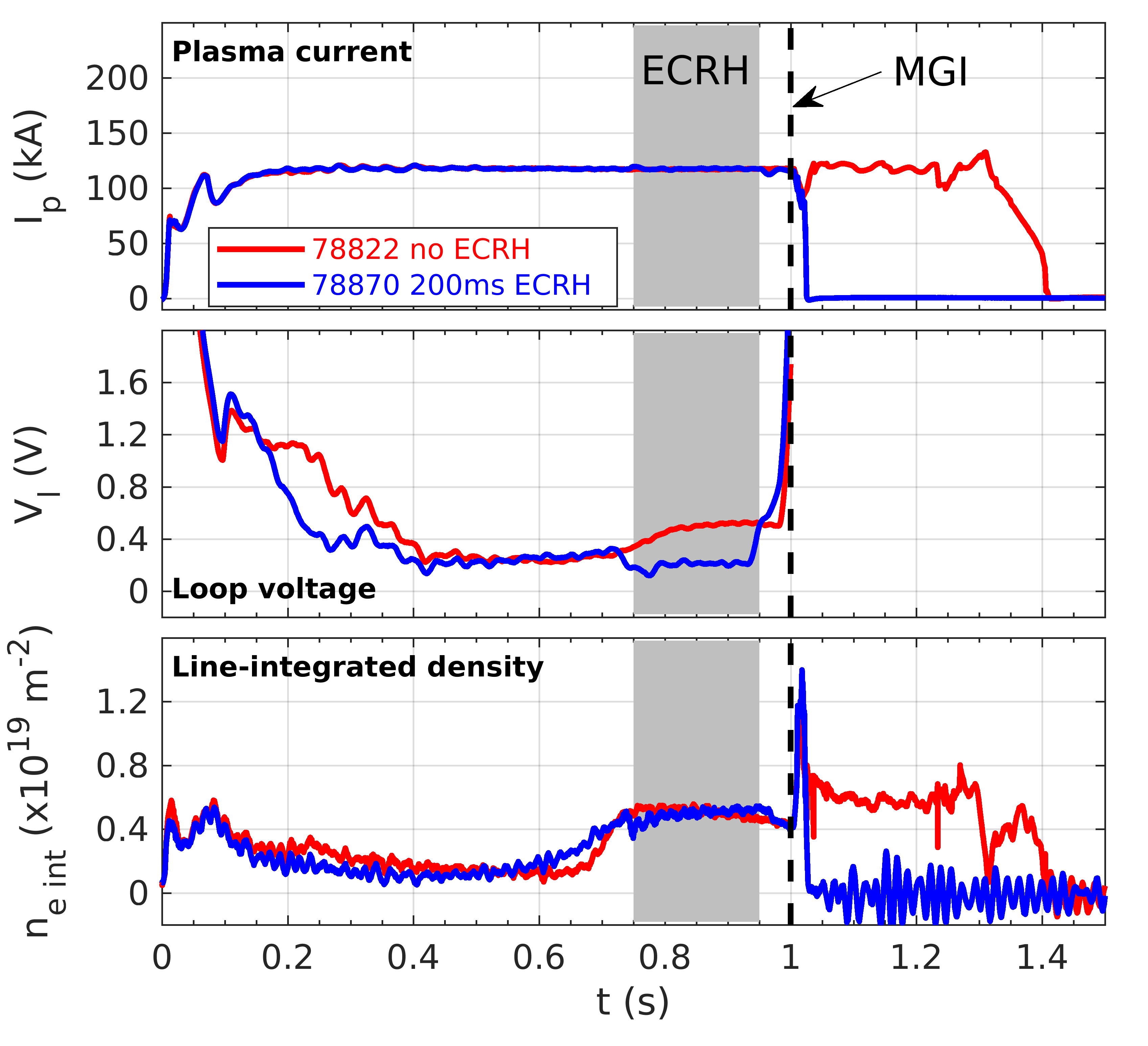}
\par\end{centering}
\begin{centering}
(a)
\par\end{centering}
\begin{centering}
\includegraphics[width=8cm]{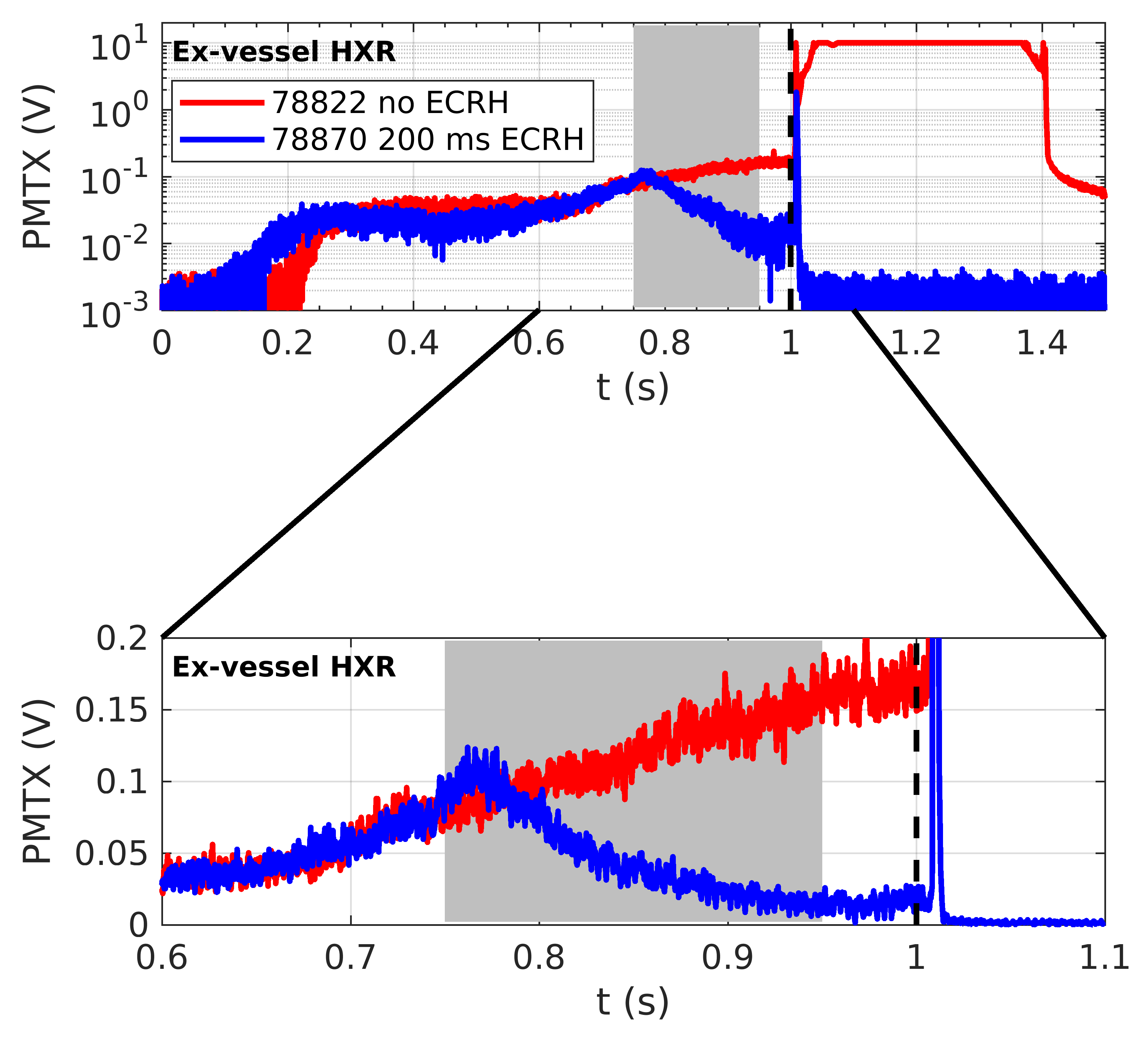}
\par\end{centering}
\begin{centering}
(b)
\par\end{centering}
\centering{}\caption{(a) Plasma current, loop voltage, line-integrated density; (b) raw
signal from the PMTX diagnostic; as a function of time for TCV discharges
\#78822 (no ECRH) and \#78870 (700 kW central ECRH between 0.75 and
0.95 s indicated by the grey area). \protect\label{fig:prevention}}
\end{figure}

TCV discharges presented in this paper feature a diverted, nearly
circular, deuterium plasma with major radius $R=0.89$ m, minor radius
$a=0.25$ m and a toroidal magnetic field on axis $B_{T}=1.44$ T.
From about $t=0.1$ s onwards, the toroidal plasma current is maintained
at $I_{p}=120$ kA through feedback control of the central solenoid
current. The presence of REs is monitored by PMTX, a photomultiplier
tube sensitive to hard X-rays (HXRs) located outside the vessel. It
is estimated that HXR photons with energies above 150 keV can traverse
the TCV chamber and be detected by PMTX. Comparisons with dosimetry
\cite{wei24} show that, in the presence of significant RE populations,
the PMTX signal remains approximately proportional to the HXR photon
energy flux. The signal scales with RE energy and RE loss rate as
HXR emission is dominated by RE-wall interaction in the absence of
high-Z impurities within the plasma \cite{hop23}.

Studies of post-disruption RE beam generation in TCV concluded that
a significant pre-MGI RE seed must exist for a RE beam to form after
a plasma discharge disruption trigerred by MGI \cite{Decker_2022}.
In TCV discharge \#78822 (Figure \ref{fig:prevention}), a strong
RE seed is generated during an initial low density phase with a line
integrated density $n_{\textrm{int}}\leq2\times10^{18}$ m$^{-2}$
between $t=0.3$ s and $0.6$ s. The HXR signal increases rapidly
until it reaches a plateau between $t=0.4$ s and $0.6$ s. As the
HXR intensity rises, the loop voltage decreases from $V_{\textrm{loop}}=1.2$
V to $0.3$ V at $t=0.4$ s. As will be demonstrated in Section \ref{sec:regimes},
this decrease in loop voltage characterises a transition from a low
to a high RE current regime, where $\sim75$\% the plasma current
is carried by REs.

In contrast with most TCV RE beam studies where the density is kept
extremely low until MGI \cite{Decker_2022,Sheikh_2024}, higher fueling
is applied in this discharge from $t=0.6$ s onward. The density rises
rapidly from $n_{\textrm{int}}=1.5\times10^{18}$ m$^{-2}$ at $t=0.6$
s to $n_{\textrm{int}}=5.3\times10^{18}$ m$^{-2}$ at $t=0.75$ s
and is then held constant until MGI at $t=1.0$ s. This density increase
is designed for RE dynamics comparison with discharge \#78870 where
ECRH power is applied and higher wall degassing inevitably increases
the density. This nearly fourfold density increase leads to a remarkably
modest reduction of the RE current, which still drives $\sim50$\%
of the plasma current at $t=0.75$ s, in accordance with an increase
in the loop voltage to $0.6$ V. The RE current remains relatively
insensitive to density variations, a characteristic property of ohmic
plasmas in the high RE current regime, as explained in Section \ref{sec:regimes}.
The HXR signal does not decrease following this significant density
increase but, instead, rises steadily. This suggests that the RE seed
energy gain -- as result of the higher loop voltage -- more than
compensates for the slightly reduced RE current. Massive injection
of neon at $t=1.0$ s ($7.2\times10^{18}$ atoms) induces a plasma
disruption. As indicated by the HXR signal saturating, a RE beam is
formed, that now drives almost all the plasma current until the shot
terminates at $t=1.4$ s. It is worth noting that, after a strong
RE seed is established, raising the electron density by a factor 4
does not prevent the formation of a RE beam following the MGI-triggered
disruption.

Figure \ref{fig:prevention} compares discharges \#78822 and \#78870,
which are very similar until t=0.75 s. In \#78870 about 700 kW of
ECRH power is applied between 0.75 and 0.95 s by an X-mode polarized
EC wave power deposited on the high-field side (HFS) at $r/a=0.25$.
With the toroidal magnetic field close to the maximum attainable at
TCV, the ECRH power is deposited as centrally as possible for a 84
GHz microwave beam launched perpendicularly to the magnetic field.
Upon ECRH application, the HXR signal increases sharply by $\sim50$\%
(Fig. \ref{fig:prevention}-b). This initial rise is followed by a
steady decay of one order of magnitude over the 200 ms ECRH phase.
Taking the HXR emission as proportional to the RE-wall collision frequency
\cite{hop23}, the initial peak indicates a sudden increase in the
RE loss rate upon ECRH onset, whereas the subsequent decay would result
from a reduction in the RE current. The steady RE depletion is explained
by an increased loss rate combined with a reduction in the RE avalanche
gain as the loop voltage decreases during ECRH. At $t=1.0$ s, an
MGI similar to that of \#78822 also triggered a disruption, but was
not followed by the formation of a RE beam. The application of ECRH
thus prevented the formation of post-disruption RE beam by expelling
the RE seed.

\begin{figure}[h]
\begin{centering}
\includegraphics[width=8cm]{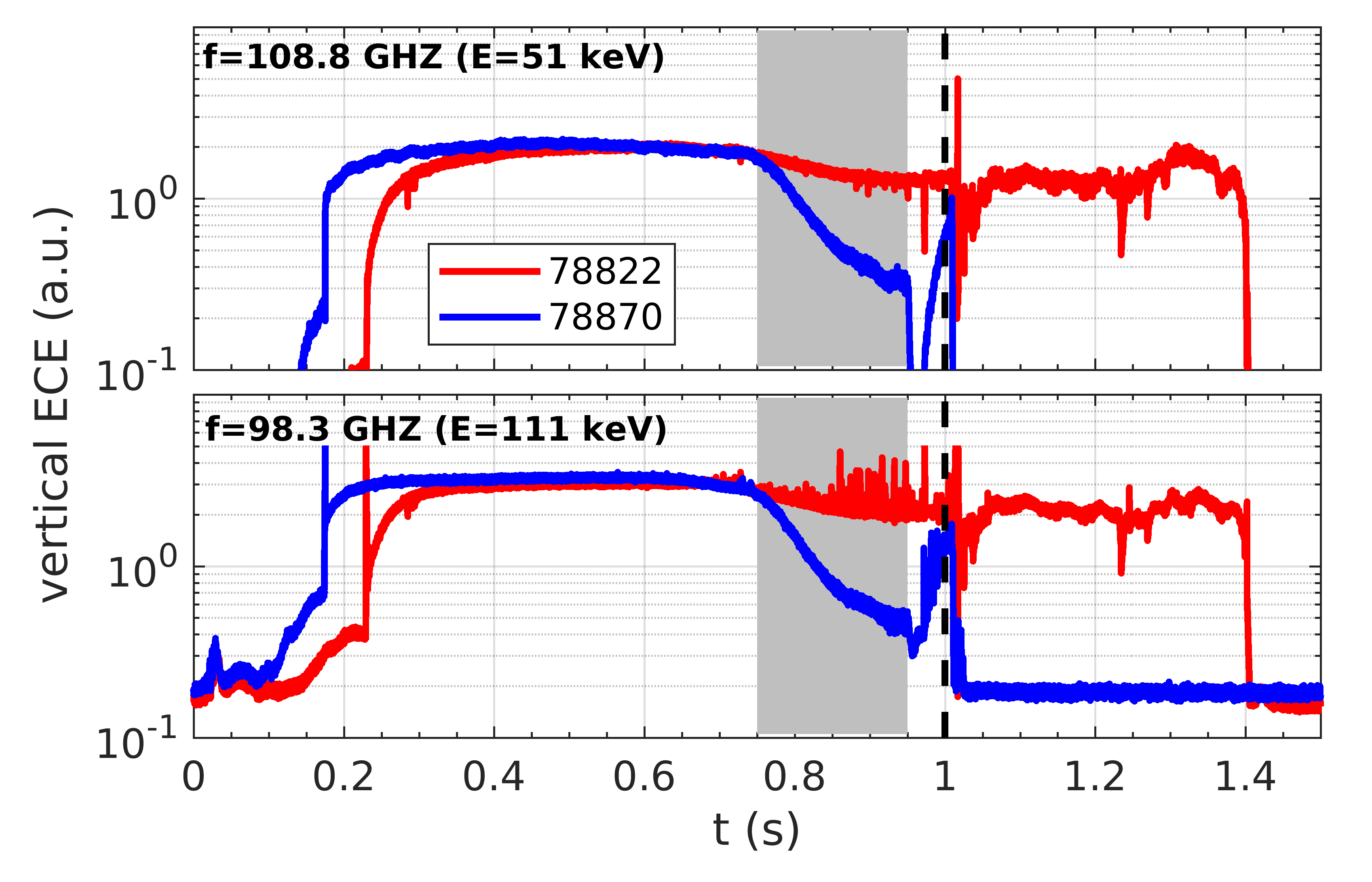}
\par\end{centering}
\centering{}\caption{Uncalibrated vertical electron cyclotron emission at 108.8 and 98.3
GHz, as a function of time for TCV discharges \#78822 (no ECRH) and
\#78870 (700 kW central ECRH between 0.75 and 0.95 s indicated by
the grey area). \protect\label{fig:vece}}
\end{figure}

Lower energy REs are also monitored using the vertical electron cyclotron
emission (VECE) diagnostic \cite{tem23}, Fig. \ref{fig:vece}. Radiation
measured at frequencies 98.3 GHz and 108.8 GHz along a vertical line
of sight intersecting the magnetic axis corresponds primarily to third
harmonic ECE emission from electrons of energy $\mathcal{E}\simeq111$
keV and $\mathcal{E}\simeq51$ keV, respectively. \textasciicircum\footnote{Higher harmonic ECE emission by higher energy electrons may not be
negligible. Estimating the importance of this contribution requires
knowledge of the electron distribution function.}At these frequencies the plasma is optically thin so VECE emission
can be taken as proportional to the integrated density of REs at the
corresponding energy along the vertical line of sight. VECE employs
a viewing dump to avoid reflections, and the density is sufficiently
low for lines of sight at the selected frequencies to originate from
the dump. The effect of spurious reflections is thus neglected \cite{tem21}.
With a critical energy $\mathcal{E}_{c}\approx50$ keV during the
ECRH phase, the 108.8 GHz radiation is emitted by electrons near critical
energy for which the electric and friction forces balance each other,
whereas the 98.3 GHz radiation is emitted by REs.

In discharge \#78822 without ECRH the VECE signals decrease by only
30\% during the higher density phase between $t=0.7$ s and $1.0$
s. In contrast, VECE signals are found to decay strongly throughout
the ECRH phase in TCV discharge \#78870, indicating a depleting RE
density at these energies. The absence of initial peak in VECE signals
upon ECRH is expected as the radiation intensity depends on the integrated
RE density along the line of sight that should not be increasing.
Conversely, this observation reinforces the hypothesis that HXR emission,
including the spike at ECRH onset, is proportional to the RE loss
rate. The decay rate from the VECE signals across the first 100 ms
in the ECRH phase is approximately 13 s$^{-1}$. It is independent
of VECE frequency and nearly identical to that of the PMTX signal
that monitors HXR emission from higher energy electrons impinging
upon the wall.

\section{Characterisation of ECRH-induced RE expulsion\protect\label{sec:power}}

\subsection{Effect of ECRH injected power on RE dynamics}

\begin{figure}
\begin{centering}
\includegraphics[width=8cm]{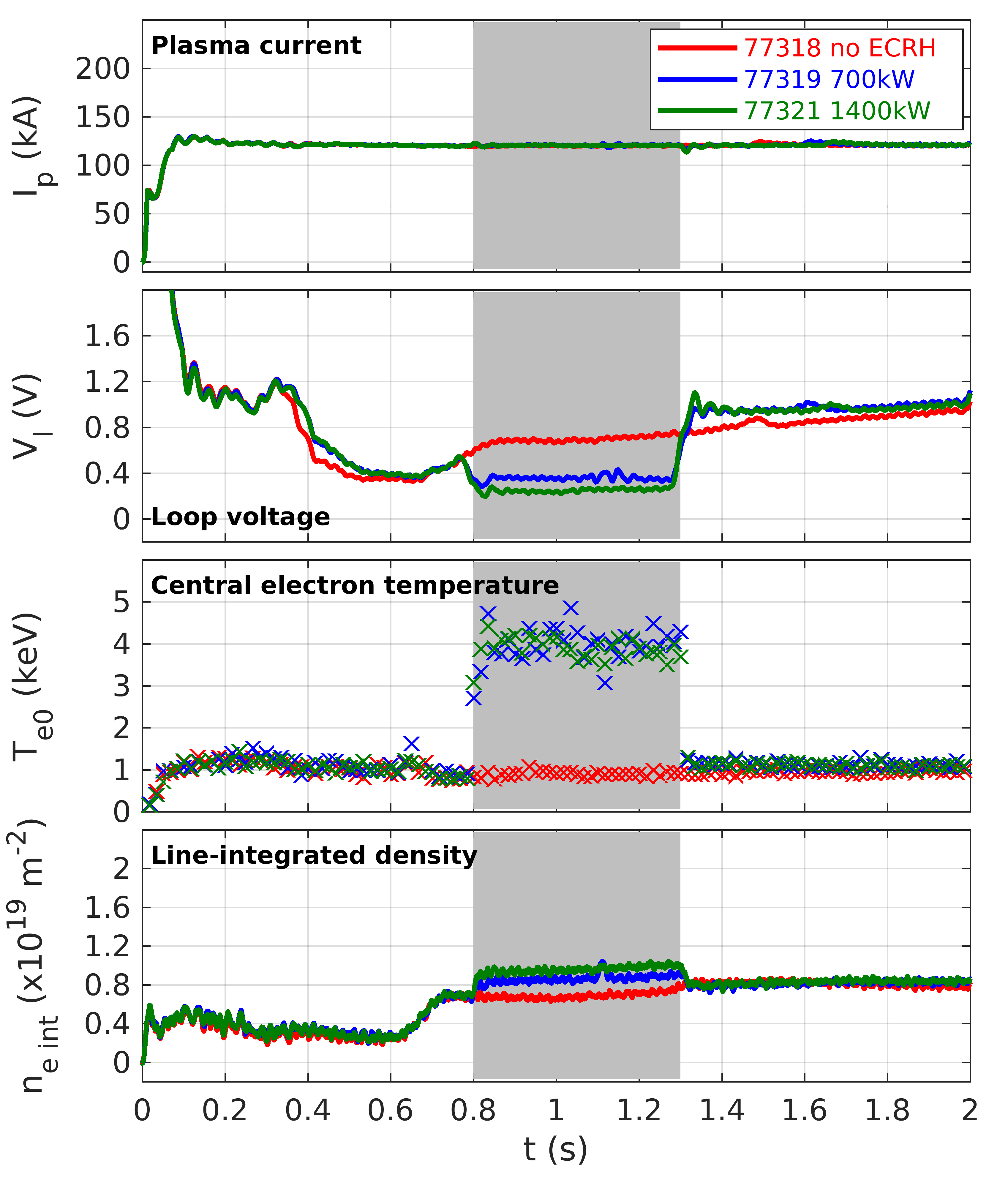}
\par\end{centering}
\begin{centering}
(a)
\par\end{centering}
\begin{centering}
\includegraphics[width=8cm]{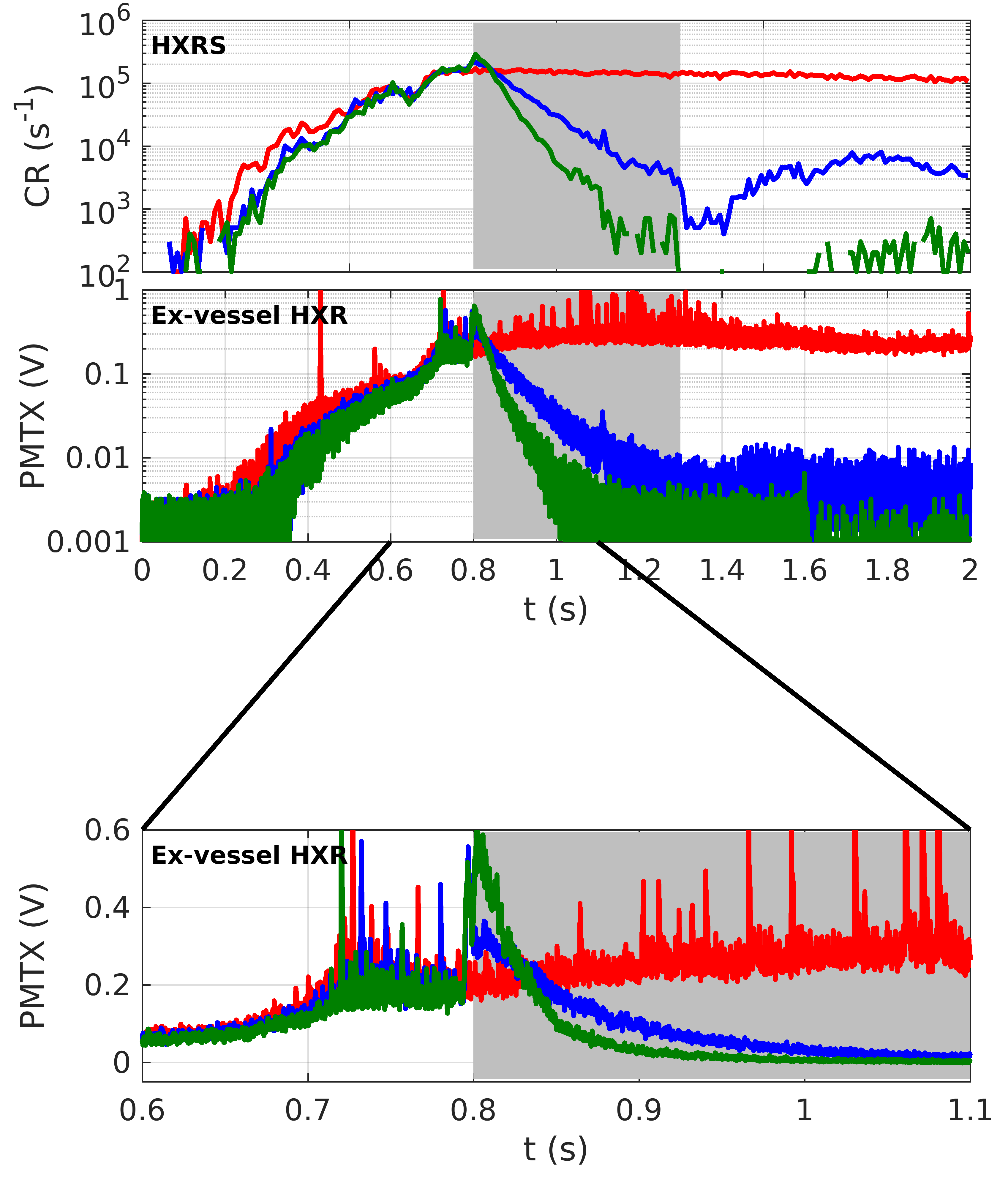}
\par\end{centering}
\begin{centering}
(b)
\par\end{centering}
\centering{}\caption{(a) plasma current, loop voltage, line-integrated density, and electron
temperature; (b) count rate from the blind detector of the top vertical
camera and raw signal from the PMTX diagnostic, as a function of time
for TCV discharges \#77318, \#77319 and \#77321 in which 0, 700 and
1400 kW of central ECRH is applied, respectively, between 0.8 and
1.3 s.\protect\label{fig:shots}}
\end{figure}

ECRH-induced RE expulsion is further probed by varying the applied
ECRH power. Fig. \ref{fig:shots} shows the time evolution of a reference
discharge without ECRH (\#77318, similar to \#77822). A significant
RE population is generated during the early, low density, phase. Around
$t=0.4$ s, the loop voltage decreases rapidly from $V_{\textrm{loop}}=1.2$
V to $0.4$ V then remains relatively steady between $t=0.45$ s and
$0.65$ s. In this high RE current regime, $\sim70$\% of the plasma
current is driven by REs, as estimated in Section \ref{sec:regimes}.
For an RE dynamics comparison between discharges, with and without
ECRH, the density is increased by $\sim3\times$ between $t=0.6$
s and $0.7$ s and then maintained.

Discharges \#77319 and \#77321 are identical to \#77318 until $t=0.8$
s, when 700kW and 1400kW of ECRH is applied, respectively, until $t=1.3$
s. The 84 GHz, X-mode polarized EC wave is launched perpendicularly
to the magnetic field and is deposited at $r/a=0.25$ on the HFS.

In addition to the PMTX diagnostic, the evolution of the RE population
is tracked by TCV\textquoteright s hard X-ray spectrometer (HXRS)
that measures the count rate from the W-shielded blind\footnote{In each HXRS camera one detector referred to as the blind detector
is shielded from the plasma by 2 cm of tungsten.} detector of the top HXRS camera \cite{gne08}, Fig. \ref{fig:shots}-b.
It is estimated that only HXR photons with energy above 500 keV can
traverse the W shielding of the camera and be registered. The HXR
signals in the ECRH-free discharge \#77318 are approximately constant
after the density plateau is reached at 0.7s. Similarly to \#78870,
the application of central ECRH in discharges \#77319 and \#77321
results in an initial peak in the HXR signals followed by a steady
decay. Similar decay rates are measured in signals by the PMTX and
HXRS diagnostics. The height of the initial HXR peak and the rate
of the subsequent decay both increase with applied ECRH power, indicating
a higher RE loss rate. In discharge \#77319 (700 kW ECRH), the PMTX
signal steadily decays during the ECRH phase, finally reaching two
orders of magnitude below its peak value. In discharge \#77321 (1400
kW ECRH), where the decay rate is higher, the PMTX signal decreases
by three orders of magnitude from the peak value, attaining a steady
state value from $t=1.1$ s onwards.

\subsection{Estimation of the RE loss rate and population reduction}

HXR emission is monitored by PMTX, located outside the vessel, and
by the HXRS camera array. It is estimated that, in the absence of
heavy impurities, this HXR emission is dominated by RE-wall interaction,
far exceeding the plasma bremsstrahlung emission \cite{hop23}, allowing
the HXR intensity to be taken as proportional to the RE loss rate.
This hypothesis is supported by the HXR intensity evolution during
ECRH as RE-plasma bremsstrahlung cannot explain an initial peak followed
by a steady decay. PMTX intensities are also found to be proportional
to radiation dosimetric measurements from the LUPIN detector \cite{wei24}.

Assuming that HXR intensity increases with the energy of the RE impinging
upon the wall and that RE losses are independent of RE energy, we
can, to first approximation, consider that the HXR intensity $\Psi$
be expressed by 
\begin{equation}
\Psi=C\nu n_{R}K(\mathcal{E})\label{eq:psi}
\end{equation}
where $\nu$ is the RE loss rate, $n_{R}$ the RE density, $K(\mathcal{E})$
is a monotonically increasing function of the average RE energy $\mathcal{E}$,
and $C$ is a time-independent constant that depends upon the RE-wall
interaction physics, the diagnostic's characteristics and the plasma
volume. The assumption that $\nu$ is indepentent of $\mathcal{E}$
is supported by the similar decay rates observed from VECE ($50<\mathcal{E}<110$
keV), PMTX ($\mathcal{E}>150$ keV) and HXRS ($\mathcal{E}>500$ keV)
signals during the ECRH phase.

As seen in Fig. \ref{fig:shots}, at the onset of ECRH, the PMTX signal
jumps, almost instantaneously, from $\Psi_{0}=C\nu_{0}n_{R0}K(\mathcal{E}_{0})$
to a peak value $\Psi_{1}=C\nu_{1}n_{R1}K(\mathcal{E}_{1})$. Assuming
that the RE density $n_{R1}\simeq n_{R0}$ and energy $\mathcal{E}_{1}\simeq\mathcal{E}_{0}$
have not changed significantly during this rapid transition, the relative
increase in loss rate can be estimated from the initial HXR rise upon
ECRH onset 
\begin{equation}
\frac{\nu_{1}}{\nu_{0}}\simeq\frac{\Psi_{1}}{\Psi_{0}}\label{eq:Rdef}
\end{equation}
From the PMTX signal in \#77319 (700 kW) and \#77321 (1400 kW) a relative
peak height $\Psi_{1}/\Psi_{0}$ -- and thus a loss rate gain $\nu_{1}/\nu_{0}$
-- of 2.0 and 3.0, respectively, are obtained.

After the initial peak, the HXR signal decreases until the end of
the ECRH phase (\#77319) or to a time-asymptotic limit (\#77321).
This is indicative of a decreasing RE density, confirmed by the absence
of any significant RE current at the end of the ECRH phase where the
loop voltage has now returned to its value before RE growth that is,
itself, consistent with neoclassical conductivity calculations \cite{sau99}.
Defining $\Psi_{2}=C\nu_{2}n_{R2}K(\mathcal{E}_{2})$ as the HXR signal
intensity at the end of the ECRH phase, across which the loss rate
$\nu_{2}\simeq\nu_{1}$ is assumed to be constant, the ratio $\Psi_{2}/\Psi_{1}$
in HXR emission provides an estimate of the corresponding relative
loss of REs 
\begin{equation}
\frac{n_{R2}}{n_{R0}}\simeq\frac{\Psi_{2}/K(\mathcal{E}_{2})}{\Psi_{1}/K(\mathcal{E}_{0})}\label{eq:Nratio1}
\end{equation}
During the ECRH phase, the parallel electric field is strongly reduced
but remains higher than the critical field ($E/E_{c}\gtrsim3$). Assuming
$\mathcal{E}_{2}\approx\mathcal{E}_{0}$ we can take
\begin{equation}
\frac{n_{R2}}{n_{R0}}\approx\frac{\Psi_{2}}{\Psi_{1}}\label{eq:Nratio2}
\end{equation}
with a value of $n_{R2}/n_{R0}\approx10^{-2}$ for \#77319 where the
RE population was still decreasing when ECRH ceased, and $n_{R2}/n_{R0}\approx10^{-3}$
for \#77321 where $\Psi_{2}$ is taken as the asymptotic limit. A
reduction of the RE density of three orders of magnitude is observed
in TCV discharge \#77321 until a steady-state is reached where residual
RE generation balances RE losses.

\begin{figure}
\begin{centering}
\includegraphics[width=8cm]{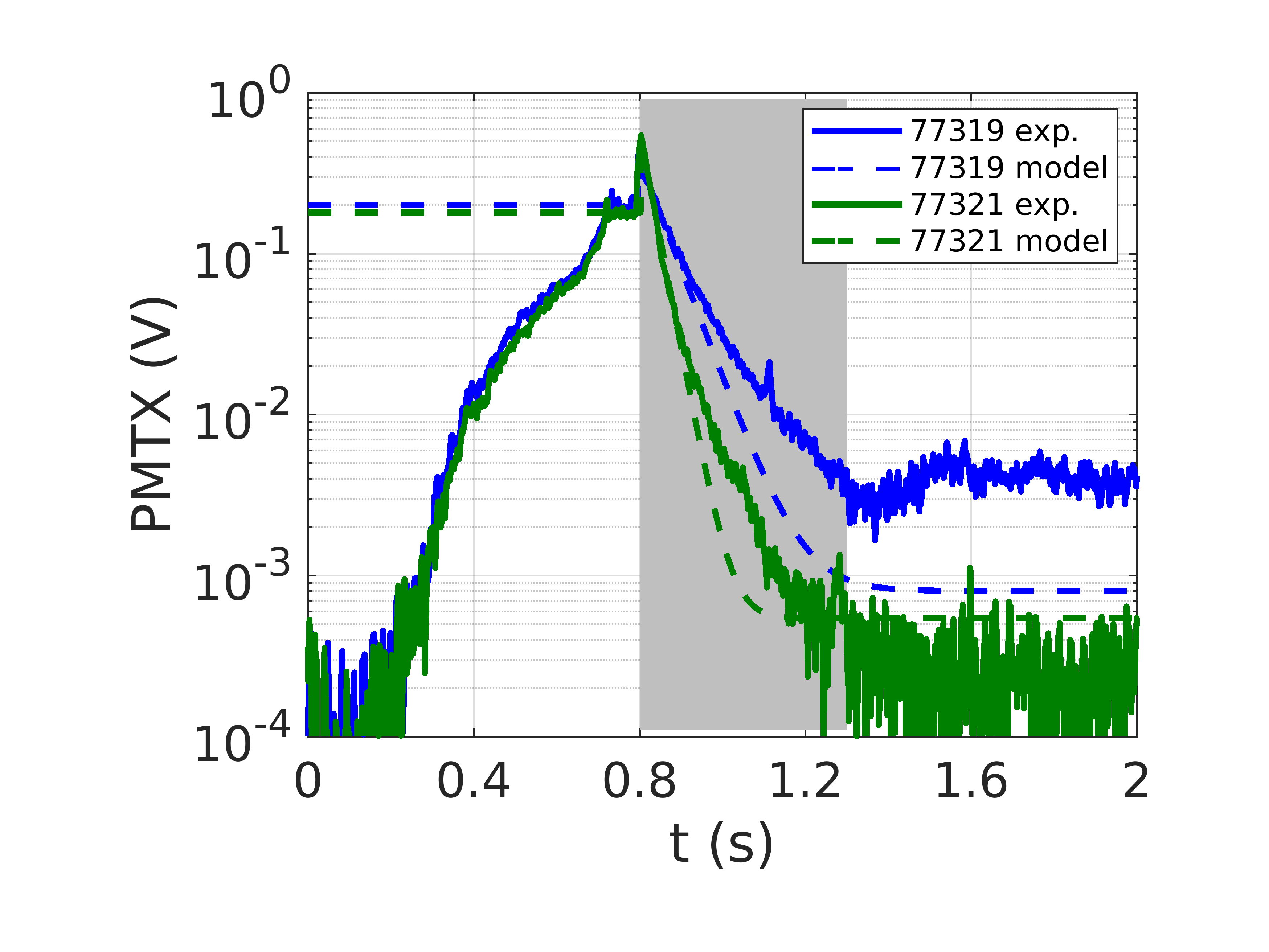}
\par\end{centering}
\begin{centering}
(a)
\par\end{centering}
\begin{centering}
\includegraphics[width=8cm]{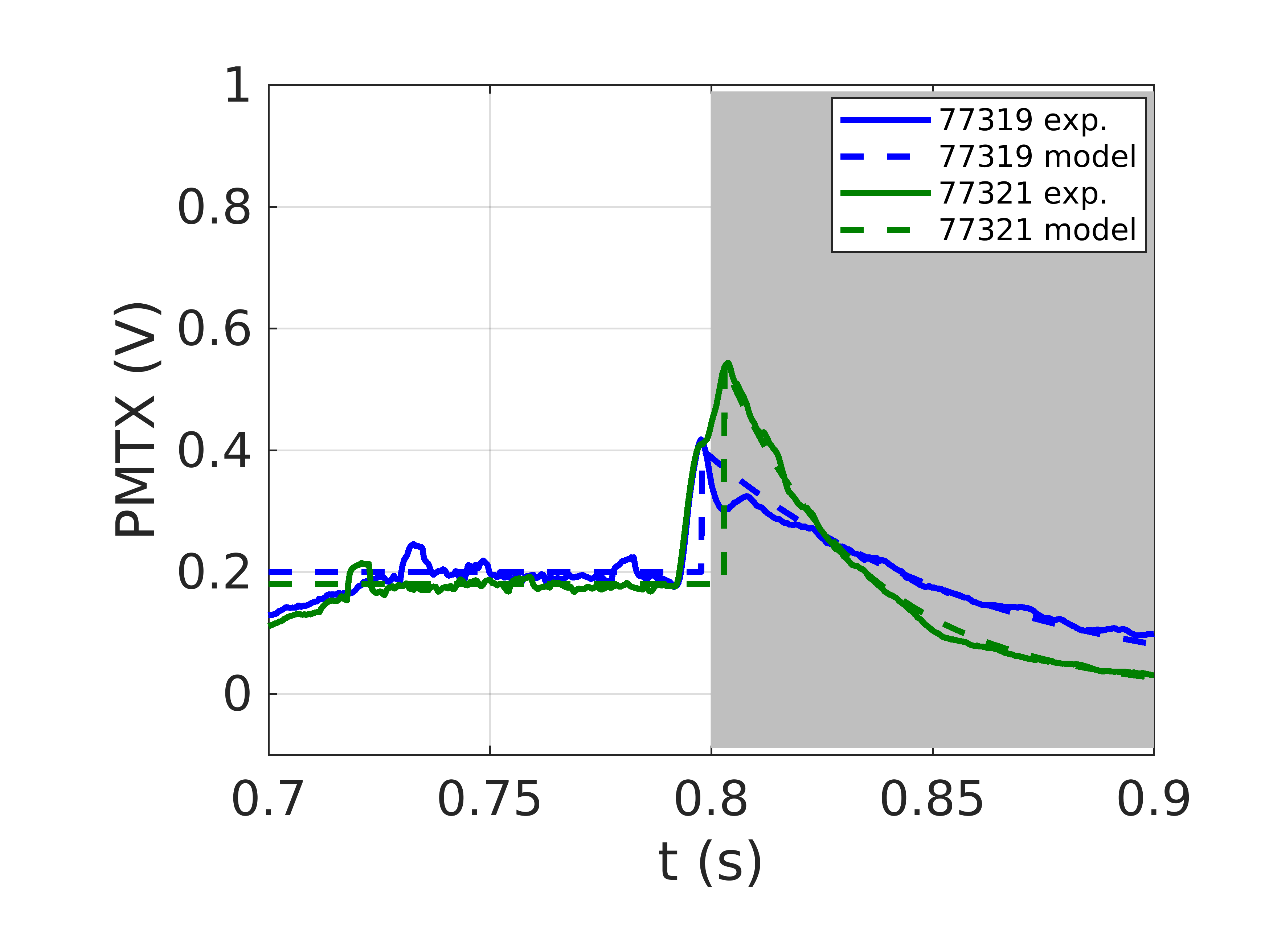}
\par\end{centering}
\begin{centering}
(b)
\par\end{centering}
\centering{}\caption{Measurement of the PMTX signal in TCV discharges \#77319 (700 kW)
and \#77321 (1400 kW) compared to the model $\Psi(t)$. \protect\label{fig:PMTX}}
\end{figure}

As described in Section \ref{sec:regimes}, the HXR intensity evolution
results from the competition between RE loss rate $\nu_{1}$ and avalanche
gain $\gamma_{A1}$. This can be estimated from the time evolution
of $\Psi$ in the early stage of the decay phase. We find $\nu_{1}-\gamma_{A1}\simeq16$
s$^{-1}$ for \#77319 (700 kW) and $\nu_{1}-\gamma_{A1}\simeq31$
s$^{-1}$ for \#77321 (1400 kW). The time evolution of the PMTX signal,
Fig \ref{fig:PMTX}, is compared to a model where $\Psi(t)=\Psi_{0}$
before ECRH ($t<0.8$ s) and $\Psi(t)=\Psi_{2}+(\Psi_{1}-\Psi_{2})\exp[-(\nu_{1}-\gamma_{A1})t]$
during ECRH ($t>0.8$ s).

In Section \ref{sec:regimes}, values of $\gamma_{A1}\simeq7$ for
\#77319 and $\gamma_{A1}\simeq4$ for \#77321, respectively, are deduced
during the ECRH phase. Combining Eq. (\ref{eq:Rdef}) with $\nu_{1}$
measurements provides an estimation for the RE loss rate, $\nu_{0}$,
before ECRH. For \#77319 (700 kW), $\nu_{1}\simeq23$ s$^{-1}$, $\Psi_{1}/\Psi_{0}\simeq2.0$
and $\nu_{0}\simeq11$ s$^{-1}$. For \#77321 (1400 kW), $\nu_{1}\simeq35$
s$^{-1}$, $\Psi_{1}/\Psi_{0}\simeq3.0$ and $\nu_{0}\simeq12$ s$^{-1}$.
Reassuringly, highly similar pre-ECRH RE loss rates are derived from
both discharges that were programmed identical until the ECRH phase.
As shown in Fig. \ref{fig:shots}, more ECRH power significantly increases
the RE loss rate. This increase, $\nu_{1}-\nu_{0}$, is found proportional
to the applied ECRH power and can be described by $(\nu_{1}-\nu_{0})/\nu_{0}\simeq P/640$
kW.

\subsection{Temperature profile and energy confinement time}

\begin{figure}
\begin{centering}
\includegraphics[width=8cm]{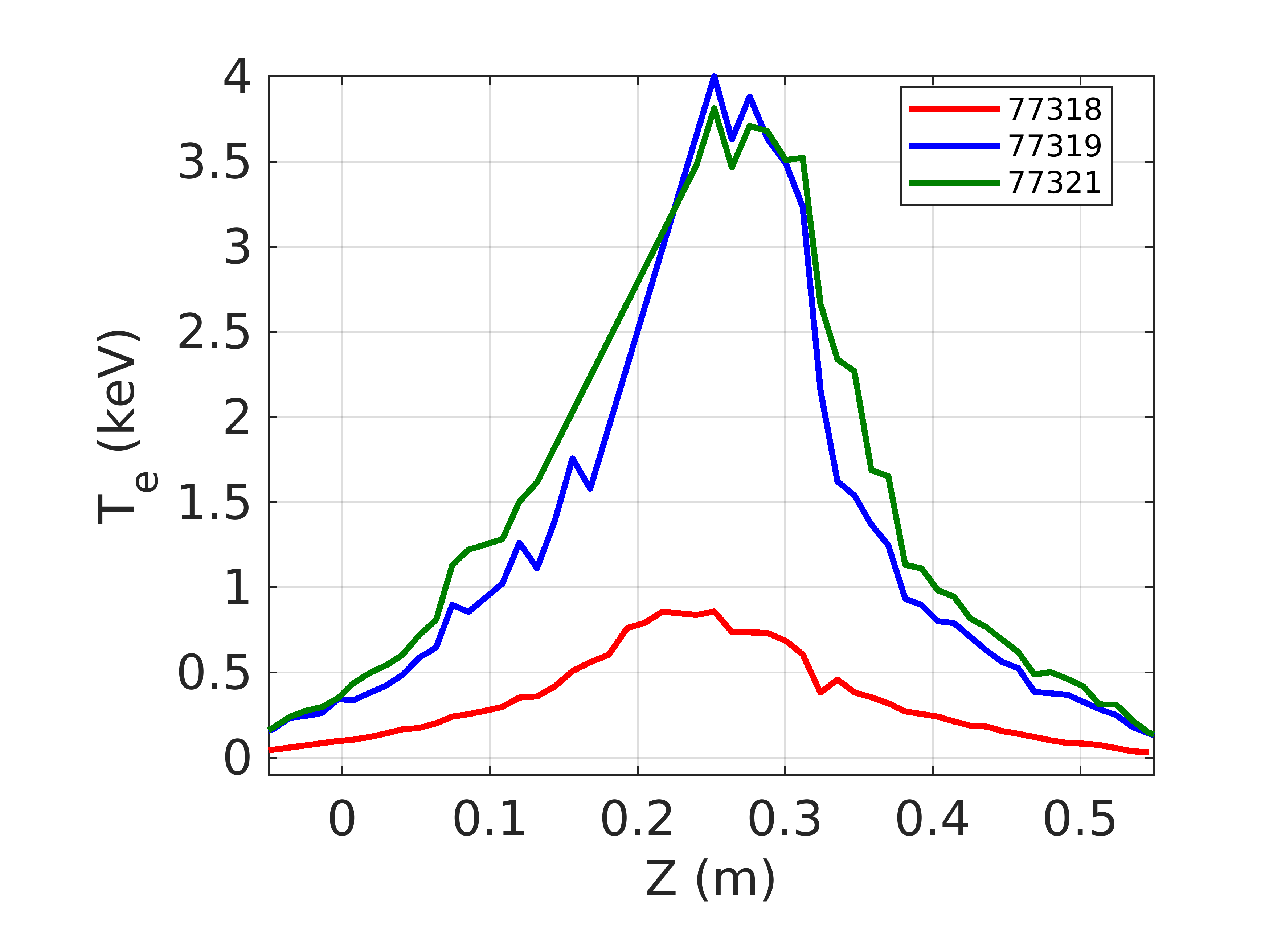}
\par\end{centering}
\centering{}\caption{electron temperature profile from Thomson Scattering (TS) measurements
averaged over the time range $t=[0.9,1.1]$ s for TCV discharges \#77318,
\#77319 and \#77321 in which 0, 700 and 1400 kW of central ECRH is
applied, respectively, between 0.8s and 1.3s. Profiles are plotted
as a function of the respective vertical position $Z$ where TS lines
of sight insersect the vertical laser path \protect\label{fig:power-te}.}
\end{figure}

The central electron temperature evolution measured by Thomson scattering
(Fig \ref{fig:shots}) is highly similar in discharges \#77319 and
\#77321, from $T_{e0}\simeq1$ keV before ECRH to $\simeq3.5$ keV
during ECRH. The electron temperature profiles measured during the
ECRH phases also overlap (Fig. \ref{fig:power-te}), suggesting that
the plasmas are in a regime of saturated micro-turbulence with stiff
electron temperature profiles where the transport is proportional
to the input power\textcolor{black}{{} \cite{gar04,gar05}}. The energy
confinement time, measured during the ECRH phase, correspondingly,
decreases from $\tau_{E1}=3.9$ ms in \#77319 (700 kW ECRH) to $\tau_{E1}=2.4$
ms in \#77321 (1400 kW ECRH) to be compared to $\tau_{E0}=9.5$ ms
in the absence of ECRH (\# 77318). The reduction in energy confinement
$\tau_{E1}^{-1}-\tau_{E0}^{-1}=150$ s$^{-1}$ for \#77319 and $\tau_{E1}^{-1}-\tau_{E0}^{-1}=310$
s$^{-1}$ for \#77321 is also found to be proportional to the applied
ECRH power. ECRH-enhanced turbulent radial transport would thus be
a possible mechanism to explain the corresponding increase in RE loss
rate \cite{hau09}.

\subsection{Effect of ECRH radial location on RE dynamics\protect\label{seq:radial}}

\begin{figure}
\begin{centering}
\includegraphics[width=8cm]{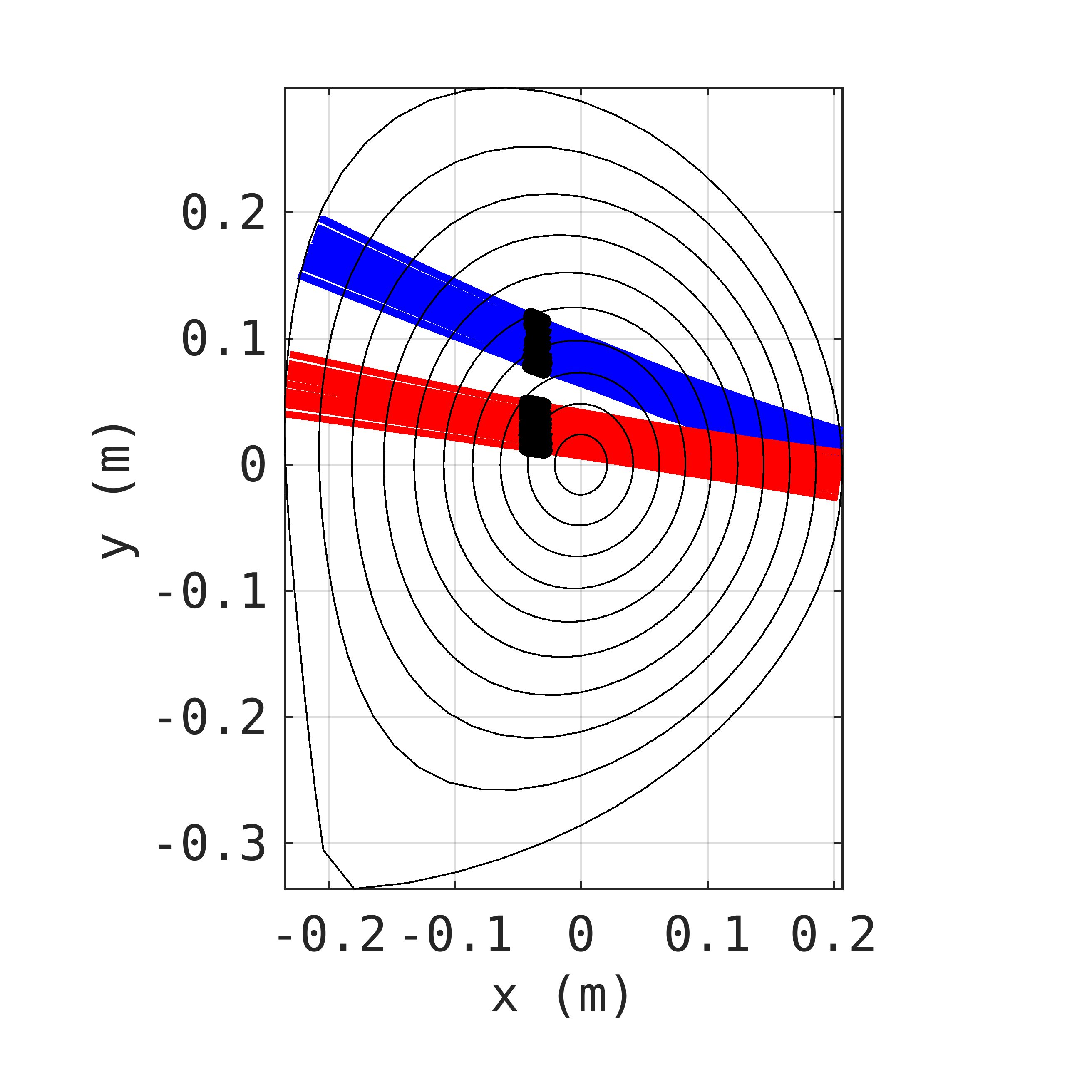}
\par\end{centering}
\begin{centering}
(a)
\par\end{centering}
\begin{centering}
\includegraphics[width=8cm]{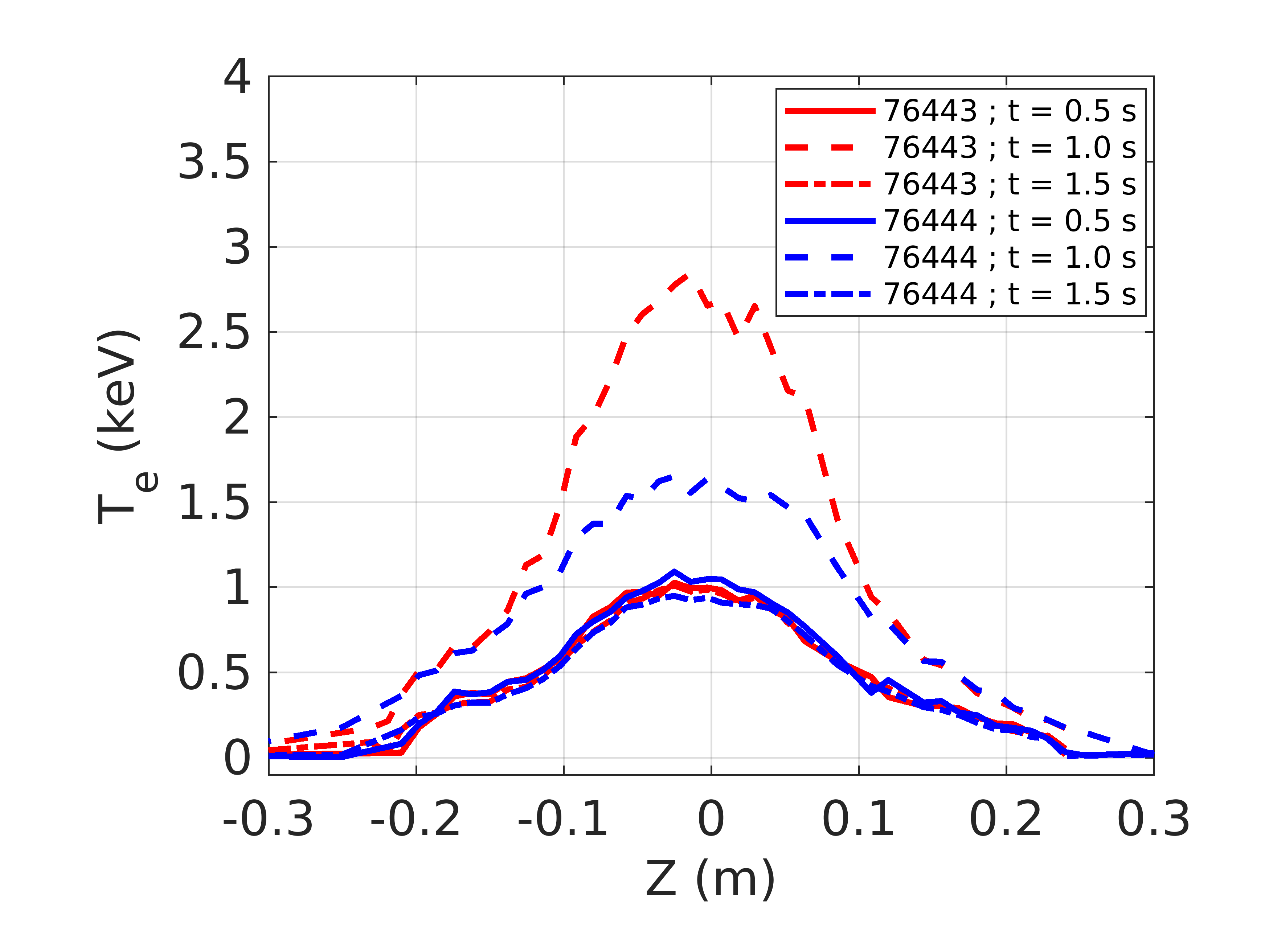}
\par\end{centering}
\begin{centering}
(b)
\par\end{centering}
\centering{}\caption{(a) ECRH propagation within a poloidal plane calculated by the RF
propagation code C3PO \cite{pey12} for TCV discharges \#77443 (red)
and \#77444 (blue) where ECRH is applied between $t=0.8$ and $1.3$
s and deposited at radial locations $r/a=0.25$ and $r/a=0.45$, respectively;
(b) correponding electron temperature profile from Thomson Scattering
measurements averaged over the time range $t=[0.4,0.6]$ s, $t=[0.9,1.1]$
s, and $t=[1.4,1.6]$ s.\protect\label{fig:prop}}
\end{figure}

\begin{figure}
\begin{centering}
\includegraphics[width=8cm]{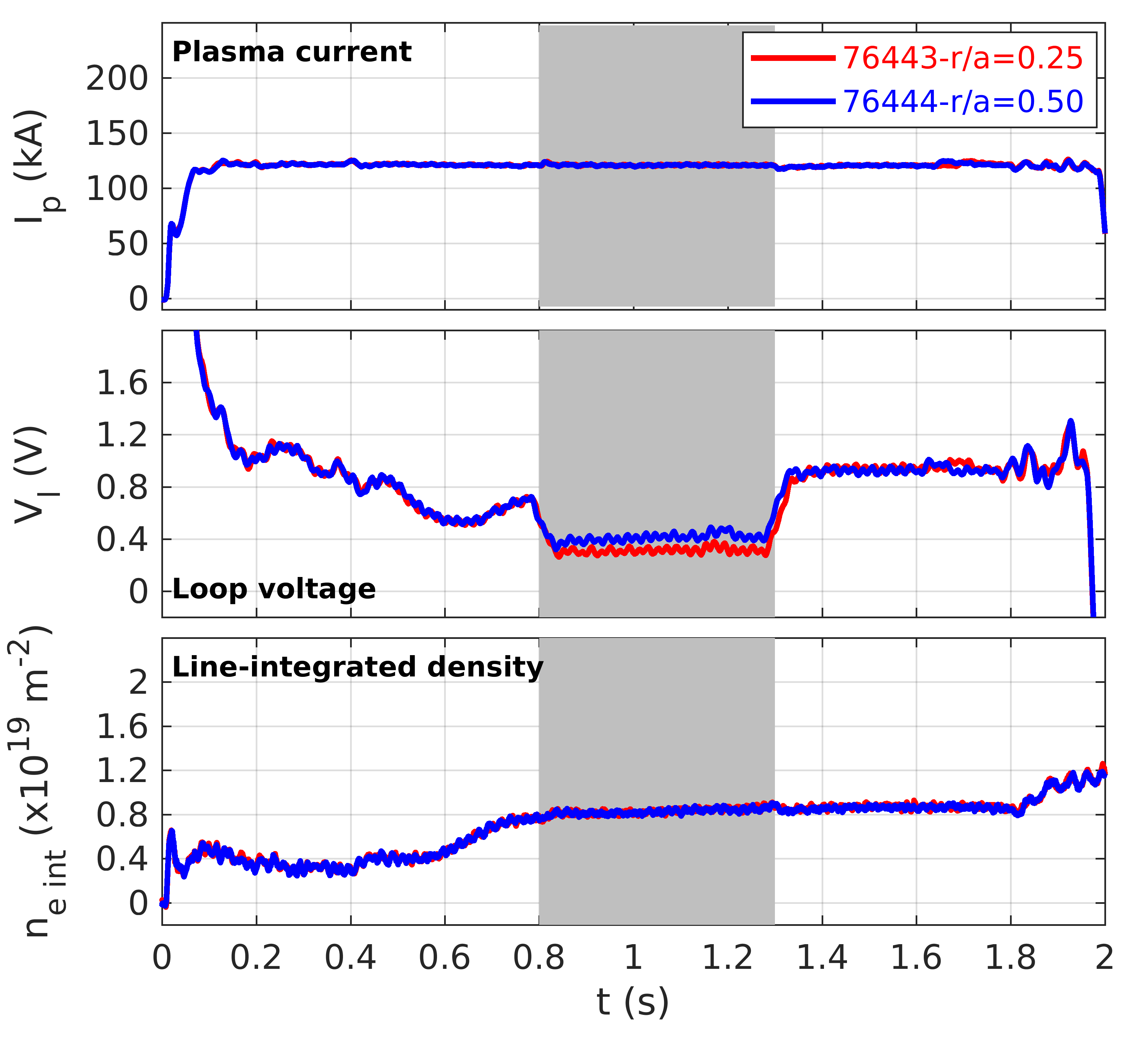}
\par\end{centering}
\begin{centering}
(a)
\par\end{centering}
\begin{centering}
\includegraphics[width=8cm]{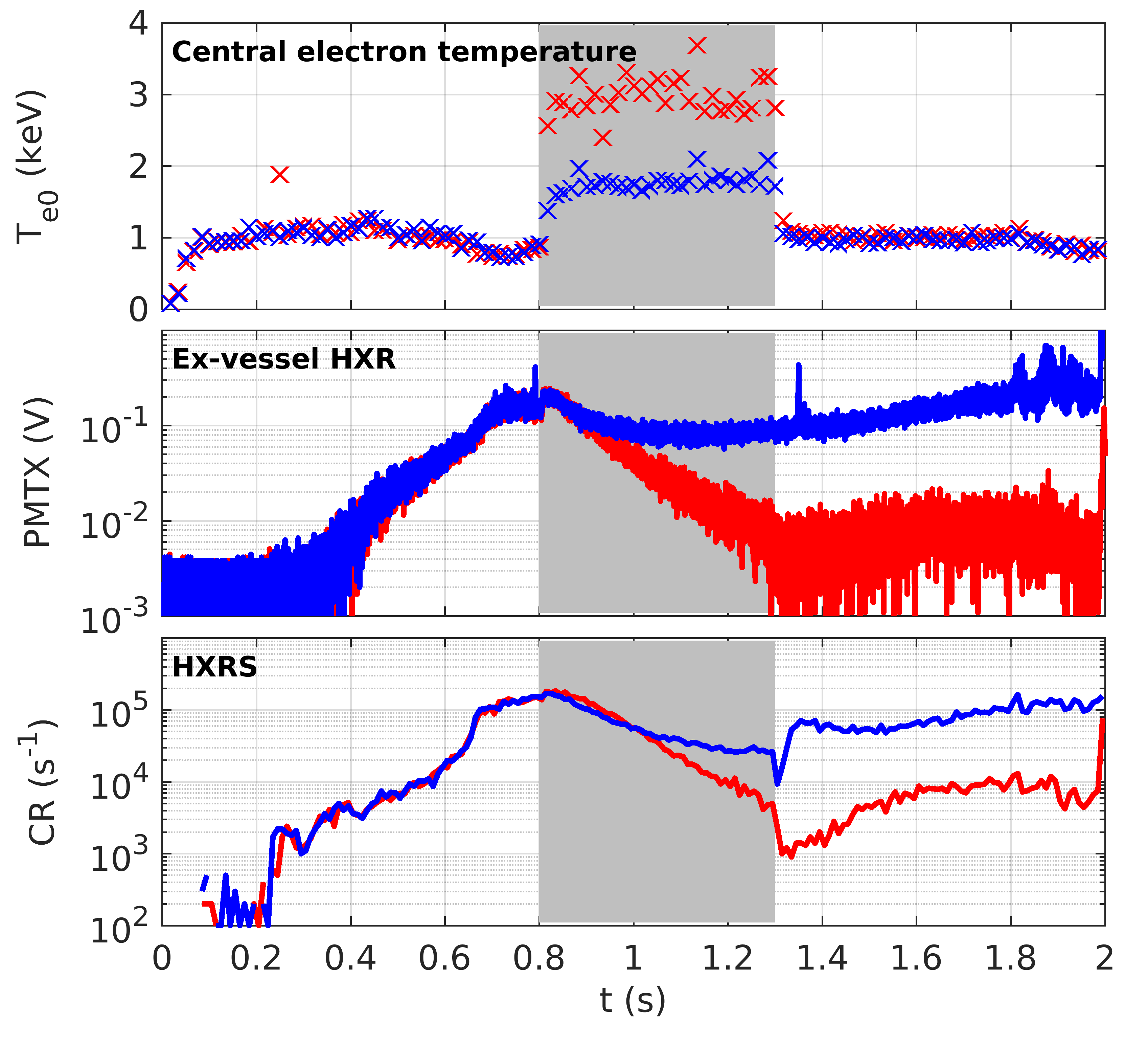}
\par\end{centering}
\begin{centering}
(b)
\par\end{centering}
\centering{}\caption{(a) Plasma current, loop voltage, line-integrated density; (b) central
electron temperature; raw signal from the PMTX diagnostic, and count
rate from the blind detector of the top vertical HXRS camera; as a
function of time for TCV discharges \#76443 and \#76444 in which 700
kW of ECRH is deposited around $r/a=0.2$ and $r/a=0.45$, respectively,
between 0.8 and 1.3 s.\protect\label{fig:radial}}
\end{figure}

In the TCV discharges presented so far, ECRH was launched perpendicularly
to the magnetic field at a poloidal angle corresponding to EC power
deposition close to the plasma centre. In a similar set of experiments,
$\sim600$ kW of ECRH (82.4 GHz, X-mode polarisation) was again launched
perpendicularly where REs carried $\sim40$\% of the plasma current.
The radial location of ECRH power deposition is varied between $r/a=0.20$
(\#76443) and $r/a=0.45$ (\#76444) by adjusting the EC poloidal launching
angle, Figure \ref{fig:prop}, with the resulting discharge time evolutions
shown in Fig. \ref{fig:radial}. Whereas the radial location of power
deposition has no significant effect on the plasma density, the increase
in central plasma temperature due to ECRH is higher for central ECRH
($T_{e0}=3$ keV) than off-axis ECRH ($1.8$ keV). The loop voltage
required to drive the plasma current is higher in discharge \#76444
($V_{\textrm{loop}}=0.4$ V) than in \#76443 ($0.3$ V). The increase
in loop voltage is, however, lower than that predicted by the effect
of electron temperature on the plasma conductivity $\sigma\sim T_{e}^{3/2}$.
Comparison with neoclassical conductivity model predictions \cite{sau99}
indicates that REs probably drive a significant proportion of the
plasma current during off-axis ECRH. This observation is confirmed
by PMTX and HXRS signals that do not decay significantly in \#76444.
Although the RE loss rate increases upon off-axis ECRH, as identified
by the initial PMTX peak, the slower decay and high asymptotic level
of HXR signals indicate that the plasma remains in the high RE current
regime described in Section \ref{sec:regimes}.

The relative height of the initial peak in PMTX signal is similar
in \#76443 and \#76444. The corresponding estimate for the increase
in RE loss rate (\ref{eq:Rdef}) is $\nu_{1}/\nu_{0}=\Psi_{1}/\Psi_{0}\simeq1.5$.
The initial decay time is also similar, with $\nu-\gamma_{A1}\simeq7$
s$^{-1}$. The HXR signal decays exponentially in discharge \#76443
with a corresponding reduction of the RE density (\ref{eq:Nratio2})
of $n_{R2}/n_{R0}\approx0.025$. For off-axis ECRH (\#76444), however,
the HXR intensity only decreases slightly before attaining a new steady-state
with a modest RE density reduction $n_{R2}/n_{R0}\approx0.5$ such
that REs still drive $\sim20$\% of the plasma current at the end
of the ECRH phase. Interestingly, this contrasted evolution of the
RE density for central vs off-axis ECRH is observed despite a similar
RE loss rate.

\section{A simple analytical model to describe ECRH-induced RE expulsion\protect\label{sec:regimes}}

\subsection{High and Low RE current regimes\protect\label{subsec:High-and-Low}}

It is shown in Sections \ref{sec:expulsion}-\ref{sec:power} that
applying central ECRH in the presence of a significant RE population
results in a strong reduction of the RE current by up to three orders
of magnitude. ECRH-induced RE expulsion can be explained using a simple
0-D model that describes how the RE population evolves from a high
RE current regime, characterised by a balance between RE losses and
avalanche generation, to a low RE current regime where RE generation
through the avalanche mechanism is negligible.

We assume a plasma with uniform electron density $n$, electron temperature
$T$, current density $J$, and parallel electric field $E$. The
evolution of the RE density $n_{R}$ is taken to be 
\begin{equation}
\frac{dn_{R}}{dt}=\Gamma-\nu n_{R}=\Gamma_{D}+\left(\gamma_{A}-\nu\right)n_{R}\label{eq:dndt}
\end{equation}
The RE generation rate $\Gamma=\Gamma_{D}+\gamma_{A}n_{R}$ includes
a Dreicer generation rate $\Gamma_{D}$ \cite{dreicer59electron}
and an avalanche growth rate $\gamma_{A}$ that can be approximated
by \cite{rosenbluth97theory}
\begin{equation}
\gamma_{A}=\frac{\alpha e}{2mc\ln\Lambda}\left(E-E_{c}\right)\label{eq:gammaA}
\end{equation}
where $\alpha$ accounts for various effects such as magnetic trapping
\cite{Nilsson_2015} that cannot be described ab-initio in a 0-D model,
and is considered here as a constant, to be determined. If $\gamma_{A}<\nu$
the RE density evolves towards a value $n_{R}=\Gamma_{D}/(\nu-\gamma_{A})$.
If $\gamma_{A}>\nu$ the RE density increases until some further saturation
mechanism comes into play. Such a mechanism may include : 1) kinetic
instabilities driven by increasing gradients in RE momentum space
\cite{Pokol_2008,Heidbrink_2019,Carnevale_2021}; 2) momentum space
vortices driven by radiation reaction forces \cite{and01,Decker_2016},
possibly enhanced by resonant mechanisms \cite{Hoppe_2020,Wijkamp_2021,Wijkamp_2024};
3) increased transport -- and corresponding loss rate $\nu$ --
resulting from changes in the plasma equilibrium due to the increasing
RE-driven current; 4) reduced Ohmic field -- with reduced avalanche
gain $\gamma_{A}$ -- as a response of the plasma control system
to the increasing RE-driven current \cite{mcd23}.

The TCV plasmas discharges described in Section 2 were under $I_{p}$-feedback
control. The loop voltage evolution shows that, here, the RE saturation
mechanism is provided by the Ohmic field reduction. We can model the
plasma current density $J$ as the sum of the thermal and RE contributions
\cite{mcd23}. Approximating the parallel RE velocity to the speed
of light
\begin{equation}
J=\sigma E+ecn_{R}\label{eq:J}
\end{equation}
where $\sigma$ is the plasma conductivity. We define $n_{Rp}\equiv J/(ec)$
as the density of REs required to drive all the plasma current. The
RE density required to drive a current $I_{p}=120$ kA in a circular
uniform plasma with $a=0.25$ m is $n_{Rp}=1.3\times10^{16}$ m$^{-3}$,
which is 2-3 orders of magnitude below the electron density of the
discharges described in this paper. The control system uses feedback
on the measured $I_{p}$ to adjust the electric field 
\begin{equation}
E=\frac{ec}{\sigma}\left(n_{Rp}-n_{R}\right)=\frac{J}{\sigma}\left(1-\frac{n_{R}}{n_{Rp}}\right)\label{eq:Eresponse}
\end{equation}
When the RE-driven current increases, the Ohmic drive is reduced and
the avalanche growth rate decreases correspondingly from its maximum
value $\gamma_{A\max}=\nu_{A}\left(1-n/n_{c}\right)$ as
\begin{equation}
\gamma_{A}=\nu_{A}\left(1-\frac{n}{n_{c}}-\frac{n_{R}}{n_{Rp}}\right)=\gamma_{A\max}-\nu_{A}\frac{n_{R}}{n_{Rp}}\label{eq:gammaA-1}
\end{equation}
where $\nu_{A}\equiv\alpha/(2\tau_{c}\ln\Lambda)$ is the maximum
avalanche growth rate in the limit $n\rightarrow0$, and $n_{c}$
is the critical density below which REs can exist
\begin{equation}
n_{c}=\frac{4\pi\varepsilon_{0}^{2}m^{2}c^{3}}{\tau_{c}e^{4}\ln\Lambda}\label{eq:nc}
\end{equation}
with $\tau_{c}=m\sigma/(e^{2}n_{Rp})$ the corresponding relativistic
collision time\footnote{Here $\tau_{c}$ is defined as a function of the expected maximum
electric field $J/\sigma$, in the absence of REs, and thus differs
slightly from its usual expression $mc/(eE)$. The interpretation
of $n_{c}$ as a critical density remains since $n_{R}\rightarrow0$
and $E\rightarrow J/\sigma$ when $n\rightarrow n_{c}$.}.

A homogeneous response of the electric field, as described by Eq.
(\ref{eq:Eresponse}), is only applicable if the RE density evolves
over time scales that are longer than the current redistribution time.
Otherwise it is necessary to solve the induction equation as performed
in codes such as GO \cite{Papp_2013} and DREAM \cite{HOPPE2021108098}.
Nonetheless, Eq. (\ref{eq:gammaA-1}) can be combined with Eq. (\ref{eq:dndt})
to analyse steady-state plasma phases where the loop voltage is uniform
and the RE-driven current is constant, i.e. $dn_{R}/dt=0$ :
\begin{equation}
\Gamma_{D}+\left[\gamma_{A\max}-\nu-\nu_{A}\frac{n_{R}}{n_{Rp}}\right]n_{R}=0\label{eq:steadystateREs}
\end{equation}
Solving Eq. (\ref{eq:steadystateREs}) provides a solution for the
steady-state RE density 
\begin{equation}
n_{R}=\frac{1}{2}\left(n_{Rh}+\sqrt{n_{Rh}^{2}-4n_{Rh}n_{Rl}}\right)\label{eq:nR}
\end{equation}
where we identify asymptotic expressions for high RE density 
\begin{equation}
n_{Rh}\equiv n_{Rp}\nu_{A}^{-1}\left(\gamma_{A\max}-\nu\right)\label{eq:HC}
\end{equation}
and low RE density 
\begin{equation}
n_{Rl}\equiv\frac{\Gamma_{D}}{\nu-\gamma_{A\max}}\label{eq:LC}
\end{equation}

In TCV plasma conditions $\Gamma_{D}\ll\nu n_{Rp}$, indicating that
the RE confinement time is too short for the Dreicer mechanism alone
to sustain a significant RE current. Here, two clearly different regimes
can be identified from Eq. (\ref{eq:steadystateREs}):
\begin{itemize}
\item a low RE current regime for $\gamma_{A\max}<\nu$ where the avalanche
gain $\gamma_{A}\simeq\gamma_{A\max}$ cannot compensate RE losses
and the steady-state RE density, $n_{R}\simeq n_{Rl}\ll n_{Rp}$,
is proportional to the Dreicer generation rate and increases with
the RE confinement time,
\item a high RE current regime for $\gamma_{A\max}>\nu$ where the avalanche
gain $\gamma_{A}\simeq\nu$ decreases from its maximum value $\gamma_{A\max}$
until it balances RE losses and the steady-state RE density is $n_{R}\simeq n_{Rh}\lesssim n_{Rp}$.
\end{itemize}
In the high RE current regime the RE density $n_{Rh}$ can be rewritten
as
\begin{equation}
n_{Rh}=\frac{1}{ec}\left[J-\sigma\left(E_{c}+E_{\nu}\right)\right]\label{eq:nrh}
\end{equation}
where $E_{\nu}\equiv2\alpha^{-1}mc\ln\Lambda\nu/e$ can be interpreted
as a correction to the critical field accounting for RE transport
and $E_{c}+E_{\nu}$ is now the effective critical field.

Exiting the high-RE current regime, or simply reducing $n_{Rh}$,
can thus be achieved by one or more of the following : a) a higher
loss rate $\nu$, b) a higher electron temperature considering $\sigma\sim T^{3/2}$,
c) a higher density as $E_{c}\sim n$, d) a lower plasma current as
$J\sim I_{p}$. REs in TCV typically require $n/n_{c}=\sigma E_{c}/J<10^{-1}$
so that, in the high RE current regime, density variations have a
weak effect on $n_{Rh}$ as observed experimentally in Section \ref{sec:power}.

Defining normalised quantities $\bar{n}\equiv n/n_{c}$, $\bar{\Gamma}_{D}\equiv\Gamma_{D}/(\nu_{A}n_{Rp})$,
and $\bar{\nu}\equiv\nu/\nu_{A}$ we can express the normalised steady-state
$\bar{n}_{R}\equiv n_{R}/n_{Rp}$ as 
\begin{equation}
\bar{n}_{R}=\frac{1}{2}\left(\bar{n}_{Rh}+\sqrt{\bar{n}_{Rh}^{2}+4\bar{\Gamma}_{D}}\right)\label{eq:nRbar}
\end{equation}
where
\begin{equation}
\bar{n}_{Rh}=1-\bar{n}-\bar{\nu}\label{eq:nRh}
\end{equation}

\begin{figure}[h]
\begin{centering}
\includegraphics[width=8cm]{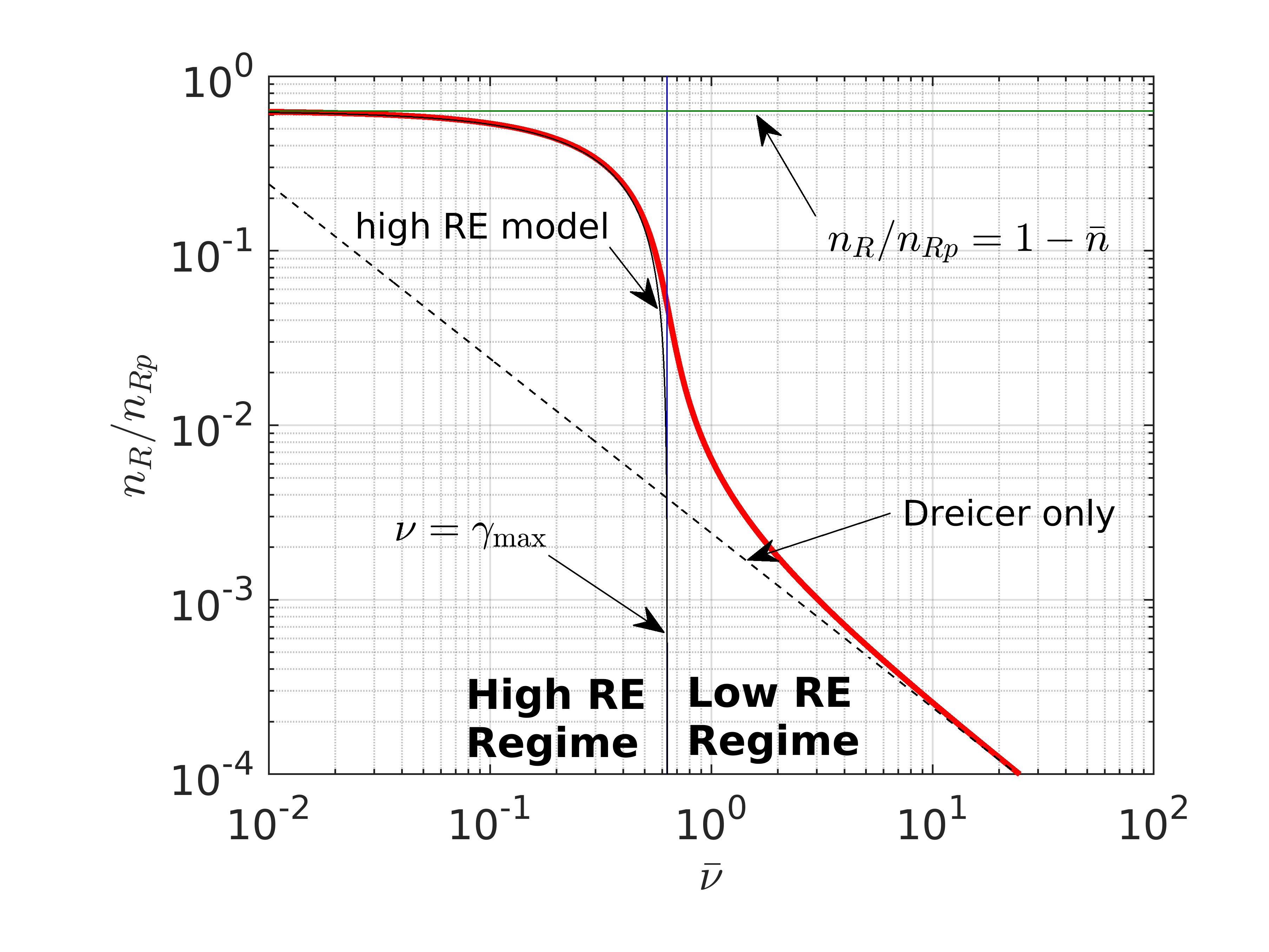}
\par\end{centering}
\centering{}\caption{\protect\label{fig:nrnu}Normalized RE density $n_{R}/n_{Rp}$ as
a function of the normalized RE loss rate $\bar{\nu}$ for parameters
$\bar{n}=0.37$, $\bar{\Gamma}_{D}=0.0024$. The asymptotic solutions
$n_{R}=n_{Rh}$ and $n_{R}=n_{Rl}$ identified for the high and low
RE regimes are indicated as dashed black lines. The transition $\nu=\gamma_{A\max}$
between the high and low RE regimes is marked as a blue vertical line.}
\end{figure}

Runaway electrons can exist only if the density is sufficiently low
so that $\bar{n}<1$, and the high RE current regime requires a low
RE loss rate $\bar{\nu}<1-\bar{n}$. When $\bar{\Gamma}_{D}\ll1$,
the high RE current regime results primarily from avalanche RE generation.
The high and low RE current regimes are illustrated in Fig. \ref{fig:nrnu}
for parameters ($\bar{n}=0.37$, $\bar{\Gamma}_{D}=0.0024$) as a
function of $\bar{\nu}$. We observe a sharp variation of the RE density
around the transition $\nu=\gamma_{A\max}$ (or equivalently $\bar{n}_{Rh}=0)$
between the high and low RE current regimes, with $n_{R}/n_{Rp}$
decreasing by two orders of magnitude for only a fourfold increase
in $\bar{\nu}$. In comparison, increasing $\bar{\nu}$ while remaining
within the high RE current regime does not affect the RE density significantly.
In the low RE current regime, increasing $\bar{\nu}$ results in the
RE density decreasing as $\bar{n}_{R}\sim1/\bar{\nu}$.

\subsection{Analysis of TCV \# 77321}

The model developed in Section \ref{subsec:High-and-Low} is applied
here to TCV discharge \#77321, described in Section \ref{sec:power}
and illustrated in Fig. \ref{fig:shots}, where five phases are identified
where $V_{\textrm{loop}}$ is approximately constant :
\begin{enumerate}
\item initial Ohmic plasma before RE growth (centered around $t=0.25$ s),
\item lowest density, low loop voltage phase ($t=0.55$ s),
\item higher density phase before ECRH ($t=0.75$ s),
\item during ECRH after the RE decay phase ($t=1.25$ s),
\item after ECRH is turned off ($t=1.50$ s).
\end{enumerate}
In each phase, the measured density and loop voltage are used to estimate
$E/E_{c}$ by assuming a uniform loop voltage. The RE loss rates are
obtained from the analysis of Section \ref{sec:power} assuming --
and verifying a posteriori -- that $\nu\gg\gamma_{A}$ during ECRH.
The main model parameters are listed in Table 1.

The Dreicer generation rate during the ECRH phase, $\Gamma_{D}/n_{Rp}=0.015$
s$^{-1}$, is derived from estimates of the steady-state RE density
and the decay rate, Eq. (\ref{eq:LC}). For comparison, a considerably
higher estimate for the Dreicer RE generation rate during the initial,
low-density, phase of the plasma discharge can be obtained from (\ref{eq:dndt}),
see \ref{sec:Dreicer}, which yields $\Gamma_{D}/n_{Rp}\approx0.2\;\textrm{s}^{-1}$.
The Dreicer growth rate is reduced by about an order of magnitude
between the low density phase and ECRH phases. Furthermore, even in
the absence of RE losses, driving a significant RE current through
the Dreicer mechanism alone would take $\sim5$ s. With the loss rate,
estimated at $\nu=10$ s$^{-1}$, the asymptotic RE density would
be limited to $n_{R}/n_{Rp}=\Gamma_{D}/(n_{Rp}\nu)\approx0.02$. For
low density Ohmic TCV plasmas in high RE current regime, RE generation
is, thus, clearly dominated by the avalanche mechanism.

\begin{table}
\begin{centering}
\begin{tabular}{|c|c|c|c|c|c|c|c|c|c|c|}
\hline 
 & $t$ & $n_{l}$ & $V_{\textrm{loop}}$ & $E/E_{c}$ & $\nu$ & $\gamma_{A}$ & $\gamma_{A\max}$ & $n_{Rh}/n_{Rp}$ & $\Gamma_{D}/n_{Rp}$ & $n_{R}/n_{Rp}$\tabularnewline
\hline 
 & s & $\times10^{19}$ m$^{-3}$ & V &  & s$^{-1}$ & s$^{-1}$ & s$^{-1}$ &  & s$^{-1}$ & \tabularnewline
\hline 
\hline 
a) & 0.25 & 0.8 & 1.00 & 27 &  & 25 & 25 &  & 0.2 & \tabularnewline
\hline 
b) & 0.55 & 0.4 & 0.40 & 22 & 10 & 10 & 25 & 0.6 &  & 0.6\tabularnewline
\hline 
c) & 0.75 & 1.3 & 0.50 & 8 & 11 & 11 & 24 & 0.5 &  & 0.5\tabularnewline
\hline 
d) & 1.25 & 2.0 & 0.25 & 3 & 35 & 4 & 4 & -4.8 & 0.015 & 0.0005\tabularnewline
\hline 
e) & 1.50 & 1.5 & 1.00 & 15 &  & 24 & 24 &  &  & \tabularnewline
\hline 
\end{tabular}
\par\end{centering}
\caption{Parameters for plasma and RE dynamics corresponding to various time
phases of TCV discharge \# 77321. The steady-state RE densities at
$t=0.55$ s and $t=0.75$ s are calculated from $I_{p}$ and $V_{\textrm{loop}}$.
An estimate for $n_{R}/n_{Rp}$ after ECRH-induced RE expulsion can
be derived from (\ref{eq:Nratio2}).}
\end{table}

\begin{figure}[h]
\begin{centering}
\includegraphics[width=8cm]{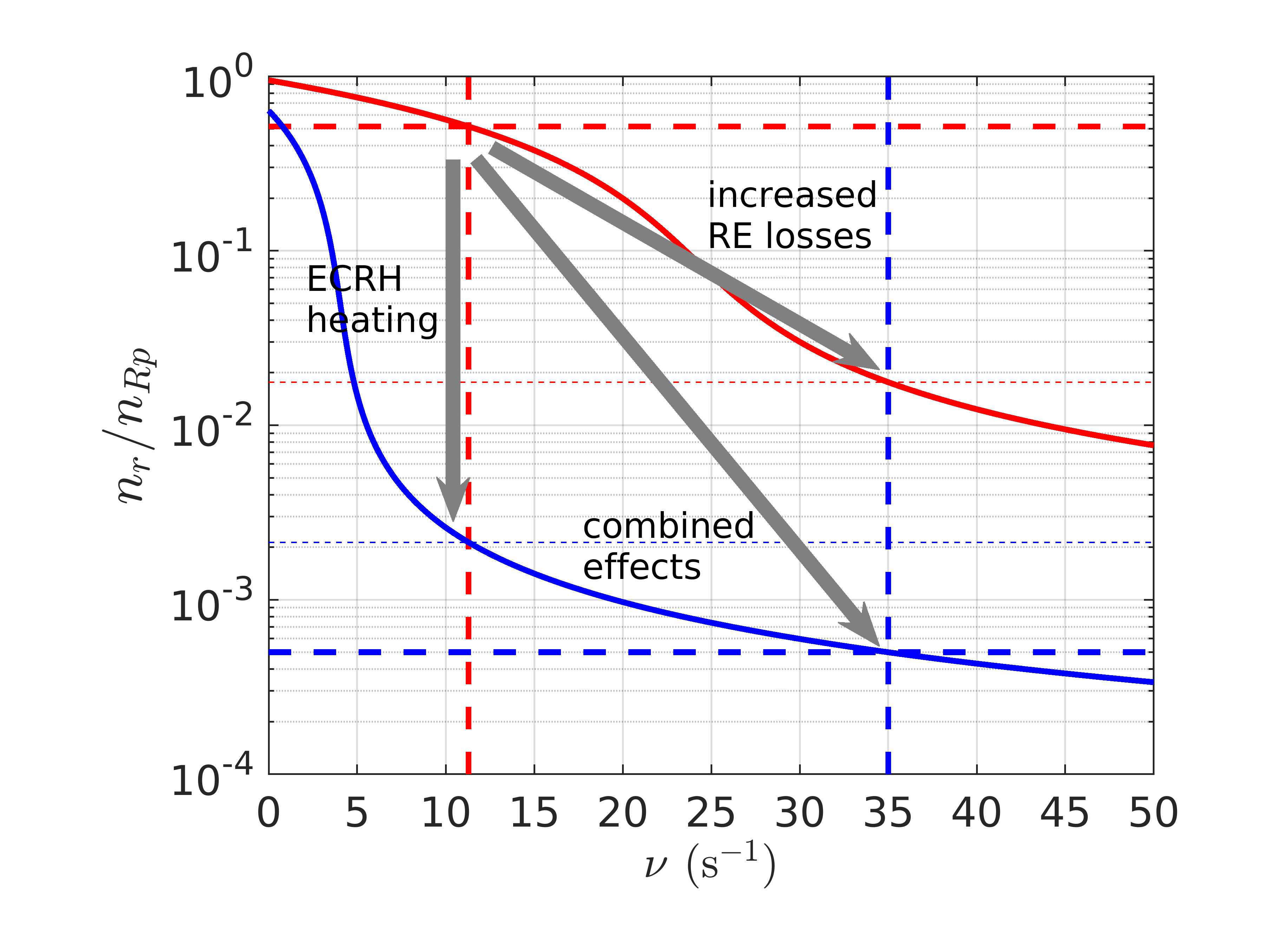}
\par\end{centering}
\centering{}\caption{\protect\label{fig:nrnu-1}Normalized RE density $n_{R}/n_{Rp}$ as
a function of the normalized RE loss rate $\bar{\nu}$ for parameters
corresponding to the pre-ECRH steady-state phase (red solid line,
$\bar{n}=0.06$, $\bar{\Gamma}_{D}=0.0078$) and the ECRH phase (blue
solid line, $\bar{n}=0.37$, $\bar{\Gamma}_{D}=0.0024$). The RE loss
rates $\nu=11$ s$^{-1}$ and $\nu=35$ s$^{-1}$ measured respectively
before and during ECRH are indicated along with the correponding $n_{R}/n_{Rp}$.}
\end{figure}

We can now describe the evolution of the RE population in discharge
\#77321 :
\begin{itemize}
\item In the initial low density phase $0.1<t<0.6$ s, $\nu<\gamma_{A\max}$,
the RE growth rate is positive and the plasma evolves towards a high-RE
current regime, indicated by the loop voltage decrease from $V_{\textrm{loop}}=1.0$
V to $0.4$ V and an ensuing steady state $\nu=\gamma_{A}$ where
avalanche generation compensates RE losses. This is a high RE current
regime.
\item From $t=0.6$ s to $t=0.7$ s the density increases from $n_{\textrm{int}}=2.5\times10^{18}$
m$^{-2}$ to $7.0\times10^{18}$ m$^{-2}$ with a consequential increase
in the critical field $E_{c}$. The loop voltage rises from $V_{\textrm{loop}}=0.4$
V to $0.5$ V to maintain the avalanche growth rate $\gamma_{A}=\nu$.
The plasma remains in a high-RE current regime.
\item During the ECRH phase, the loss rate increases. In addition, ECRH
heating increases the conductivity of thermal electrons and the control
system reacts by reducing the loop voltage. The reduced avalanche
RE generation can no longer compensate RE losses. The RE population
decays to a balance between Dreicer generation and RE losses. The
plasma is now in a low RE current regime.
\item After ECRH, the loop voltage is higher than in the pre-ECRH phase.
However, with a density now considerably higher than for the $t<0.6$
s phase, Dreicer generation is comparatively low and RE regeneration
over the remaining duration of the discharge can be neglected.
\end{itemize}
The transition from a high RE current regime before ECRH to a low
RE current regime during ECRH is illustrated graphically for discharge
\#77321 by plotting the normalised RE density $n_{R}/n_{Rp}$ as a
function of the absolute RE loss rate $\nu$ (Fig. \ref{fig:nrnu-1}).
We observe a three orders of magnitude decrease in the RE density
from the combined effects of increased RE losses from $\nu=11$ s$^{-1}$
to $35$ s$^{-1}$ and the increased conductivity of thermal electrons
attributed to ECRH (transition from the red to the blue curve).

We can infer that the heating effect of ECRH -- and corresponding
increase in conductivity -- would have led to a RE density reduction
by a factor $\sim100$ without an increase of the RE loss rate. Conversely,
the increase in RE loss rate alone would have led to a RE density
reduction by a factor $\sim20$. The combined effects induce a decrease
of RE density by a factor $\sim1000$ as observed experimentally.

\subsection{Analysis of ECRH phase in EC power and radial location scans}

\begin{table}
\begin{centering}
\begin{tabular}{|c|c|c|c|c|c|c|c|c|c|c|c|}
\hline 
shot \# & $P_{EC}$ & $r/a$ & $t$ & $n_{l}$ & $V_{\textrm{loop}}$ & $E/E_{c}$ & $\nu$ & $\gamma_{A}$ & $\gamma_{A\max}$ & $n_{Rh}/n_{Rp}$ & $n_{R}/n_{Rp}$\tabularnewline
\hline 
 & kW &  & s & $\times10^{19}$ m$^{-3}$ & V &  & s$^{-1}$ & s$^{-1}$ & s$^{-1}$ &  & \tabularnewline
\hline 
\hline 
77321 & 1400 & 0.25 & 1.25 & 2.0 & 0.25 & 3 & 35 & 4 & 4 & -4.8 & 0.0005\tabularnewline
\hline 
77319 & 700 & 0.25 & 1.25 & 1.8 & 0.35 & 4 & 23 & 7 & 7 & -1.8 & (0.001)\tabularnewline
\hline 
76443 & 600 & 0.20 & 1.25 & 1.6 & 0.3 & 4 & 13 & 6 & 6 & -0.9 & (0.002)\tabularnewline
\hline 
76444 & 600 & 0.45 & 1.25 & 1.6 & 0.45 & 6 & 16 & 9 & 9 & -0.5 & (0.003)\tabularnewline
\hline 
\end{tabular}
\par\end{centering}
\caption{Parameters for plasma and RE dynamics corresponding to the ECRH phase
of discharges described in Section \ref{sec:power}. An approximate
potential asymptotic post-expulsion RE density $n_{R}/n_{Rp}$ is
evaluated in shots \#77319, \#76443 and \#76444 by using the value
$\Gamma_{D}/n_{Rp}=0.015$ s$^{-1}$ calculated in \#77321.\protect\label{tab:scans}}
\end{table}

The model developed in Section \ref{subsec:High-and-Low} is applied
to the ECRH power and radial deposition scans by calculating $n_{Rh}$
during the ECRH phase. Results are summarized in Table \ref{tab:scans}.
For TCV discharge \#77319 (700 kW), illustrated in Fig. \ref{fig:shots},
it is found that $\nu=23$ s$^{-1}$ and $\gamma_{A\max}=7$ s$^{-1}$
-- to compare with $\nu=35$ s$^{-1}$ and $\gamma_{A\max}=4$ s$^{-1}$
for TCV discharge \#77321 (1400 kW). Despite a slightly higher loop
voltage and a lower RE loss rate, the ECRH phase of \#77319 is clearly
in the low RE regime, indicated by the steady decay. A Dreicer rate
similar to that of \#77321 would indicate an asymptotic RE density
of $n_{R}/n_{Rp}=0.001$, which is compatible with experimental observations
where the decaying RE density attained $n_{R}/n_{Rp}=0.005$ at the
end of the ECRH phase.

A similar observation can be made for shot \#76443 (600 kW, $r/a=0.20$),
where the ECRH phase characterised by $\nu=13$ s$^{-1}$ and $\gamma_{A\max}=6$
s$^{-1}$ is also in the low RE current regime and the RE density
measured at the end of the ECRH phase $n_{R}/n_{Rp}=0.01$ is above
the estimated asymptotic value $n_{R}/n_{Rp}=0.002$.

In the case of off-axis ECRH in TCV, \#76444 (600 kW, $r/a=0.45$),
this 0-D analysis yields $\nu=16$ s$^{-1}$ and $\gamma_{A\max}=9$
s$^{-1}$ again indicating a low RE current regime. The predicted
asymptotic RE density $n_{R}/n_{Rp}=0.003$ is two orders of magnitude
below the estimated steady-state value $n_{R}/n_{Rp}=0.2$ deduced
from experimental measurements. The difference could indicate that
the 0-D model is no longer applicable in the case of off-axis ECRH.
A possible explanation is that off-axis ECRH increases the transport
of REs located closer to the plasma edge but does not significantly
affect core RE confinement.

\section{Discussion and conclusions}

Runaway electrons are observed in TCV discharges when the parallel
electric field exceeds the Connor-Hastie critical field by an order
of magnitude or more. For very low density plasmas, the RE growth
rates are sufficiently strong for REs to drive a significant fraction
of the plasma current through an avalanche-dominated RE generation
regime. The plasma current feedback system reacts to the increasing
RE-driven current by reducing the parallel electric field until avalanche
RE generation and RE losses balance. Here, the RE density is predicted
to be resilient to density variations, as observed experimentally.
Applying central ECRH to a TCV plasma in the high-RE current regime
was seen to strongly reduce the RE density by up to three orders of
magnitude. Such a reduction is sufficient to prevent the formation
of a RE beam following a disruption triggered by massive gas injection.
It is shown that the strongly reduced RE population is associated
with a transition to a low-RE current regime where avalanche RE generation
cannot compensate for RE losses. A 0-D model, developed in Section
\ref{sec:regimes}, demonstrates that this transition results from
the combined effects of an increased RE loss rate and a reduced Ohmic
electric field -- with a correspondingly reduced RE avalanche generation
rate -- due to the increased conductivity of bulk electrons heated
by ECRH. The RE decay rate and RE density reduction factor both increase
with ECRH power. A radial scan of the ECRH deposition location showed
that RE expulsion is much stronger for central than off-axis ECRH.
A plausible explanation is that core RE confinement remains mostly
unaffected by off-axis ECRH.

In a further set of experiments, not reported herein, varying the
ECRH toroidal angle between discharges did not produce a noticeably
different result. The HXR signal decay and asymptotic level are found
to be similar for perpendicular ECRH, co, and counter-current drive
configurations. It is worth noting that, under these conditions, the
plasma current driven by EC waves was not significant and the loop
voltage measured in the co- and counter-ECCD cases was near identical.
It remains interesting to determine the effect of EC toroidal angle
on RE expulsion in situations where the EC-driven current is significant.
More generally, a strong reduction in the RE population resulting
from a transition from a high- to low- RE density regime can be expected
employing other sources of additional heating and current drive to
reduce the Ohmic field, or by applying any method that augments RE
losses.

An increased RE loss rate during ECRH is demonstrated experimentally
but an underlying physics mechanism has not been identified. Resonant
EC interaction is not expected between REs and EC waves and no significant
MHD activity is observed during the ECRH phase. Kinetic instabilities
may develop during ECRH in the presence of REs as observed for post-disruption
RE beams in FTU \cite{Carnevale_2021}. The increased RE loss rate
could alternatively be associated with an increase in microturbulence
with ECRH power \cite{hau09}. We observe that the electron temperature
measured for 700 kW and 1400 kW of ECRH is nearly identical. This
clear stiffness in the temperature profile suggests that turbulence
driven radial energy transport increases with ECRH power. It is also
plausible that RE transport increases with core turbulence as suggested
by other fast electron experiments and simulations \cite{Cazabonne_2023}.
Future experiments are envisioned to characterise the effect of ECRH
on turbulent transport to determine whether it can be more directly
correlated with RE losses.

In a reactor-scale tokamak, RE generation during disruptions is expected
to be dominated by the avalanche process, which depends upon the magnitude
of the RE seed. ECRH-induced expulsion of any RE seed may be a path
to prevent or limit RE beam formation should a disruption occur later
in the discharge. Applying ECRH early in the discharge could reduce
startup RE generation and prevent a possible transition to the dangerous
slideaway regime. The simplified 0-D model presented in this paper
can be calibrated to track the low and high RE regimes identified
herein and provide a real-time control system where appropriate remedial
action can be taken earlier and/or after a RE seed population has
been formed.

In the TCV experiments described herein, RE expulsion during ECRH
is particularly strong as the combination of increased RE transport
and reduced avalanche RE generation is sufficient to initiate a transition
from high- to low RE current regime, thereby decreasing the RE density
by orders of magnitude. Additional work is required to determine the
effect of ECRH on an existing RE seed for larger tokamaks.

\appendix

\section{Estimation of the initial Dreicer RE generation \protect\label{sec:Dreicer}}

\begin{figure}[h]
\begin{centering}
\includegraphics[width=8cm]{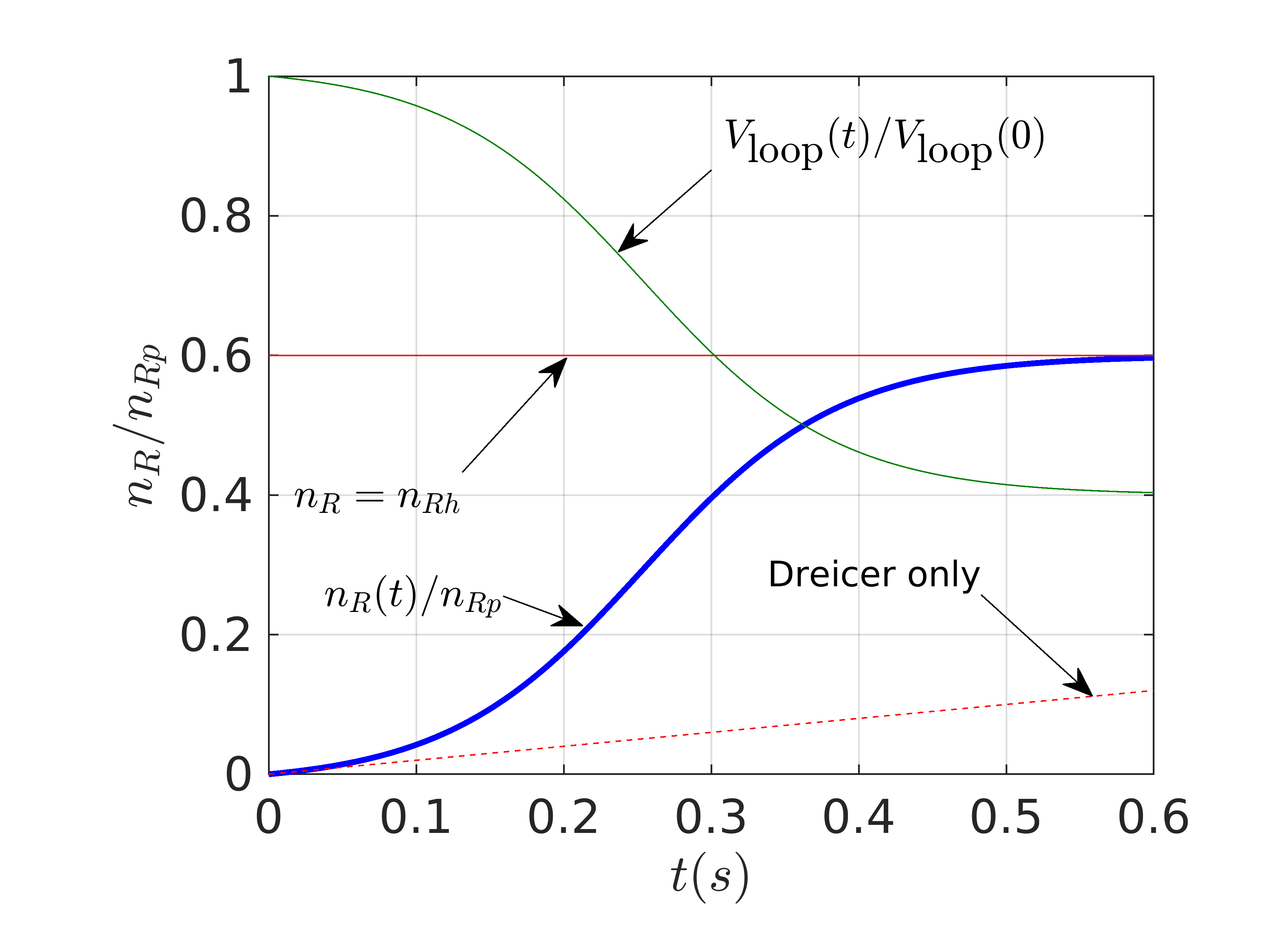}
\par\end{centering}
\centering{}\caption{\protect\label{fig:nrnu-1-1}Time evolution of the normalized RE density
$n_{R}/n_{Rp}$ (solid blue line) for parameters corresponding to
the initial low density phase of TCV discharge \#77321 ($\Gamma_{D}/n_{Rp}=0.2\;\textrm{s}^{-1}$,
$\gamma_{A\textrm{res}}\approx15$ s$^{-1}$, $n_{Rh}/n_{Rp}=0.6$).
The solid red line indicates the time-asymptotic limit $n_{R}=n_{Rh}$
and the dashed red line shows the RE density evolution accounting
only for Dreicer generation $n_{R}=\Gamma_{D}t$. The corresponding
relative evolution of $V_{\textrm{loop}}$ is also indicated (solid
green line).\protect\label{fig:tevol}}
\end{figure}

RE growth during the initial low density phase of discharges described
in this paper is analysed by calculating the time-dependent RE density
combining equations (\ref{eq:dndt}) and (\ref{eq:gammaA-1})

\begin{equation}
\frac{dn_{R}}{dt}=\Gamma_{D}+\gamma_{A\textrm{res}}\left[1-\frac{n_{R}}{n_{Rh}}\right]n_{R}\label{eq:evol-1}
\end{equation}
where we define the resulting growth rate $\gamma_{A\textrm{res}}\equiv\gamma_{A\max}-\nu$
and the asymptotic high-RE solutions is $n_{Rh}=n_{Rp}\gamma_{A\textrm{res}}/\nu_{A}$.
Starting from $n_{R}=0$, the solution of Eq. (\ref{eq:evol-1}) is
approximately
\begin{equation}
\frac{n_{R}}{n_{Rh}}\simeq\Gamma_{D}\frac{1-\exp\left[-\gamma_{A\textrm{res}}t\right]}{\Gamma_{D}+\gamma_{A\textrm{res}}n_{Rh}\exp\left[-\gamma_{A\textrm{res}}t\right]}\label{eq:timesol}
\end{equation}
where we assume $\Gamma_{D}\ll\gamma_{A\textrm{res}}n_{Rh}$ . We
note that $n_{R}(0)=0$ and $n_{R}(\infty)=n_{Rh}$ . The corresponding
time evolution is illustrated in Fig. \ref{fig:tevol} for parameters
corresponding to the initial low density phase in TCV discharge \#77321
described in Section \ref{sec:power}. An estimate for the Dreicer
rate is obtained by measuring the elapsed time $t_{\textrm{infl.}}$
to the inflexion point $d^{2}n_{R}/dt^{2}=0$ for which
\begin{equation}
\Gamma_{D}=\gamma_{A\textrm{res}}n_{Rh}\exp\left[-\gamma_{A\textrm{res}}t_{\textrm{infl.}}\right]\label{eq:gammaDbar}
\end{equation}
In the initial low density phase the inflexion point $t_{\textrm{infl.}}$
can be estimated to be about 0.25 s. With $\gamma_{A\textrm{res}}\approx15$
s$^{-1}$ and $n_{Rh}/n_{Rp}\approx0.6$ we find
\begin{equation}
\frac{\Gamma_{D}}{n_{Rp}}\approx0.2\;\textrm{s}^{-1}\label{eq:gammadHC}
\end{equation}

\ack{}{}

The authors wish to thank A. Battey and H. Choudhury for insightful
conversation on ECRH-induced RE expulsion mechanisms. This work has
been carried out within the framework of the EUROfusion Consortium,
partially funded by the European Union via the Euratom Research and
Training Programme (Grant Agreement No 101052200 -- EUROfusion).
The Swiss contribution to this work has been funded by the Swiss State
Secretariat for Education, Research and Innovation (SERI). Views and
opinions expressed are however those of the author(s) only and do
not necessarily reflect those of the European Union, the European
Commission or SERI. Neither the European Union nor the European Commission
nor SERI can be held responsible for them.

\section*{References}{\bibliographystyle{unsrt}
\bibliography{references}
}
\end{document}